\documentclass[twocolumn,xcolor]{aastex631}

\newcommand\ec{$\eta$~Car}
\newcommand\hst{{\it HST}}
\newcommand\stis{{\it STIS}}
\newcommand\cmfgen{{\it CMFGEN}}
\newcommand\kms{km~s$^{-1}$}

\newcommand\Vinf{V$_{\infty}$}
\newcommand{\Rsun}{\hbox{$R_\odot$}}
\newcommand{\Lsun}{\hbox{$L_\odot$}}
\newcommand{\Msun}{\hbox{$M_\odot$}}
\newcommand{\Mdot}{\hbox{$\dot M$}}
\newcommand{\Msunyr}{\hbox{$M_\odot\,$yr$^{-1}$}}
\newcommand{\tauv}[1]{\hbox{\small ($\tau=#1$)}}

\newcommand{\red}{\color{red}}

\received{tbd}
\revised{tbd}
\accepted{tbd}

\submitjournal{ApJ}

\shorttitle{\ec: evolving view}
\shortauthors{Gull et al.}

\begin{document}
\title{Eta Carinae: an evolving view of the central binary, its interacting winds and its foreground ejecta}

\correspondingauthor{Theodore R. Gull}
\author[0000-0002-6851-5380]{Theodore R. Gull}
\affiliation{Exoplanets \&\ Stellar Astrophysics Laboratory, NASA/Goddard Space Flight Center, Greenbelt, MD 20771, USA}\affiliation{Space Telescope Science Institute, 3700 San Martin Drive. Baltimore, MD 21218, USA}
\author[0000-0001-5094-8017] {D. John Hillier}
\affiliation{Department of Physics \&\ Astronomy \& Pittsburgh Particle Physics,
    Astrophysics, \&\ Cosmology Center (PITT PACC),\hfill\\  University of Pittsburgh,  3941 O'Hara Street, Pittsburgh, PA 15260, USA}
\author[0000-0001-9853-2555]{Henrik Hartman}
\affiliation{Materials Science \&\ Applied Mathematics, Malm\"{o}
University, SE-20506 Malm\"{o}, Sweden}

\author[0000-0002-7762-3172]{Michael F. Corcoran}
\affiliation{CRESST \&\ X-ray Astrophysics Laboratory, NASA/Goddard Space Flight Center, Greenbelt, MD
20771, USA}
\affiliation{The Catholic University of America, 620 Michigan Ave., N.E. Washington, DC 20064, USA}
\author[0000-0002-7978-2994]{Augusto Damineli}
\affiliation{Instituto de Astronomia, Geof\'isica e Ci\^encias Atmosf\'ericas, Universidade de S\~ao Paulo, Rua do Mat\~ao 1226, Cidade Universit\'aria S\~ao Paulo-SP, 05508-090, Brasil}
\author[0000-0003-2971-0439]{David Espinoza-Galeas}\affiliation{The Catholic University of America, 620 Michigan Ave., N.E. Washington, DC 20064, USA}
\affiliation{Departamento de Astronomia y Astrofisica, Facultad de Ciencias Espaciales, Universidad Nacional Autonoma de Honduras, Bulevar
Suyapa, Tegucigalpa, M.D.C, Honduras, Centroamerica}
\author[0000-0001-7515-2779]{Kenji Hamaguchi}
\affiliation{CRESST \&\ X-ray Astrophysics Laboratory, NASA/Goddard Space Flight Center, Greenbelt, MD 20771, USA}
\affiliation{Department of Physics, University of Maryland Baltimore County, 1000 Hilltop Circle, Baltimore, MD 21250, USA}
\author[0000-0002-0284-0578]{Felipe Navarete}
\affiliation{SOAR Telescope/NSF's NOIRLab, Avda Juan Cisternas 1500, 1700000, La Serena, Chile}
\author[0000-0003-2636-7663]{Krister Nielsen}
\affiliation{The Catholic University of America, 620 Michigan Ave., N.E. Washington, DC 20064, USA}
\author[0000-0001-7697-2955]{Thomas Madura}\affiliation{Department of Physics \&\ Astronomy, San Jose State University, One Washington Square, San Jose, CA 95192, USA}
\author[0000-0002-4333-9755]{Anthony F. J. Moffat}{\affiliation{D$\acute{e}$pt. de physique, Univ. de Montr$\acute{e}$al, C.P. 6128, Succ. C-V, Montr$\acute{e}$al, QC H3C 3J7, Canada \& Centre de Recherche en Astrophysique du Qu$\acute{e}$bec, Canada}}
\author[0000-0002-5186-4381]{Patrick Morris}\affiliation{California Institute of Technology, IPAC, M/C 100-22, Pasadena, CA 91125, USA}
\author[0000-0002-2806-9339]{Noel D. Richardson}
\affiliation{Department of Physics \&\ Astronomy, Embry-Riddle Aeronautical University, 3700 Willow Creek Rd, Prescott, AZ 86301, USA}
\author[0000-0002-9213-0763]{Christopher M. P. Russell}\affiliation{Department of Physics and Astronomy, Bartol Research Institute, University of Delaware, Newark, DE 19716, USA}
\author[0000-0001-7673-4340]{Ian R. Stevens}\affiliation{School of Physics \&\ Astronomy, University of Birmingham, Birmingham
B15 2TT, UK}
\author[0000-0001-9754-2233]{Gerd Weigelt}
\affiliation{Max Planck Institute for Radio Astronomy, Auf dem H\"ugel 69, 53121 Bonn, Germany}

 
\begin{abstract}
 FUV spectra of \ec, recorded across two decades with  \hst/STIS, document multiple changes in resonant lines caused by dissipating extinction in our line of sight. The FUV flux has increased nearly ten-fold which has led to increased ionization of the multiple shells within the Homunculus and photo-destruction of H$_2$. Comparison of observed resonant line profiles with \cmfgen\ model profiles allows separation of wind-wind collision and  shell absorptions from the primary wind P\,Cygni profiles. The dissipating occulter preferentially obscured the central binary and interacting winds relative to the very extended primary wind. We are now able to monitor changes in the colliding winds with orbital phase. High velocity transient absorptions occurred across the most recent periastron passage,  indicating acceleration of the primary wind by the secondary wind which leads to a downstream, high-velocity bowshock that is newly generated every orbital period. There is no evidence  of changes in the properties of the binary winds.

\end{abstract}



\section{Introduction}\label{sec:intro}
Eta Carinae (\ec) is the most luminous, most massive star within 3 kpc. In the 1840s it underwent a massive explosion, ejecting a nebula now known as the
Homunculus, which has expansion velocities of $\sim$650\,\kms \citep{Hillier92a,Steffen14}. The mass of the nebula is believed to be from 10 to  greater than 45\,\Msun\ \citep{Morris17}. An additional, less energetic eruption occurred in the 1890's \citep{SmithFrew11}. Possible earlier eruptions are suggested by  \cite{Kiminki16}.
These events, and possibly others, have given rise to a very complex circumstellar environment that is associated with \ec.

\ec\ provides an important, nearby example of those rare binary  stars which inhabit the upper region of the  Hertzsprung-Russell diagram, and which are responsible for extreme supernovae, gamma-ray bursts and black holes near the upper boundary of the lower black hole mass gap \citep{2021arXiv211103634T}. Despite more than two  centuries of dedicated study, \ec’s evolutionary path and its current evolutionary state are not  constrained. Periodic spectral \citep{Damineli96,Damineli97} and X-ray emission variations \citep{Ishibashi99a, Corcoran05a} show that \ec\ is a long-period, highly-eccentric colliding-wind binary system consisting of a near 100 solar mass primary, \ec-A, and a smaller, massive-star companion, \ec-B.  We don’t know whether the presence of the companion star is fundamental to the current state of the primary (through mass and/or angular momentum exchange, for example), or incidental to it.  It’s very difficult to constrain the stellar parameters of the companion since it is lost in the glare of the primary star at nearly  all wavelengths.  
 
The clearest evidence we have of the companion is from the X-rays produced when the companion’s wind, deduced to be moving near 3000 \kms\ based on analysis of X-ray spectra \citep{Pittard02}, collides with the dense, slow-moving wind (420 \kms, \cite{Groh12a}) of the primary star.   Indirect evidence comes from transient high-velocity absorptions \citep{Groh10,Richardson16,Gull21a} extending to speeds of 900 to 2200 \kms\ (two to five times the terminal speed of the primary wind), presumably arising from the primary wind being accelerated by the faster wind of the secondary.  Variable ionization of the circumstellar ejecta due to a  combination of the UV radiation from the companion star and from the primary star (which leaks through the colliding wind “shock cone” which surrounds the secondary, periodically illuminating different circumstellar structures like a searchlight) also provides important information about the temperature and luminosity of the companion star.  Analysis of the changing circumnebular ionization across periastron passages and over longer timescales indirectly suggests that the companion is a hot star, about a third as massive as \ec-A and probably about one fifth as luminous \citep{Verner05a, Mehner10} — similar to other extreme stars in the Carina Nebula.  
 
Our view of the central binary has been obscured by the presence of an optically thick “occulter” \citep{Hillier92a,Weigelt95,Falcke96,Hillier01,Hillier06} close to the star and in our line of sight (LOS).  Evidence shows that this occulter has now  dissipated  to such an extent that \ec\ is once again near naked-eye visibility for the first time since before the Great Eruption (when it was near a visual magnitude of $\approx$\,3.5) \citep{Damineli19,Damineli21}.  The decline in this occulter provides us a much improved view of the FUV (1150 to 1680\AA) emission from the inner binary. Absorptions at velocities near $-$513 \kms\ arise from the 1840s eruption which produced the Homunculus, while slower absorptions near $-$146 \kms\ arise from the 1890s event. Many weaker absorptions at velocities ranging from $-$121 to $-$1000 \kms\ are also detectable \citep{Gull06}. Seeing-limited spectra include emission-line contributions from the Weigelt clumps \citep{Weigelt86} and fainter structures at $-$40 \kms\ \citep{Gull16}. 

In this paper we examine changes in  FUV spectra of \ec, recorded  over the past 20 years with the Hubble Space Telescope/ Space Telescope Imaging Spectrograph (\hst/STIS)\footnote{Based on observations made with the NASA/ESA Hubble Space Telescope, obtained at the Space Telescope Science Institute,\\ which is operated by the Association of Universities for Research in Astronomy, Inc., under NASA contract NAS5-26555.\\ The new observations were  part of {\it Chandra} programs 11200668 and 13200857 and are associated with \hst\ programs 12013 and 12750.}. We consider spectra taken at intervals across nearly four orbital cycles with many recorded at nearly the same orbital phase, necessary to disentangle non-phase related (secular) variations from phase-dependent ones.

The primary objectives of our work are to search for direct evidence
of the companion star and its wind, to search for signatures of wind-wind interaction, and to quantify and understand the changes in the spectrum as a function of orbital phase and over time. Of particular interest is whether changes in the spectrum over the last two decades, sampling three binary orbits, require changes in the primary star, or whether they can be explained by solely by variations in the circumstellar medium around \ec.

This paper is organized as follows:
\begin{itemize}
\item Section \ref{sec:stage} describes the various structures  known to lie in our LOS and potential contributions to the resonant line absorptions.
\item Section \ref{sec:OBS} catalogues the observations discussed herein. 
\item Section \ref{sec:resonant}  presents profiles of resonant lines of interest broken down into these subsections: 
\begin{itemize}
\item wind lines of interest (\ref{sec:windlines}), 
\item   a description of the \cmfgen\ model used to help disentangle ``classical'' wind features  (\ref{sec:CMFGENmod}).
\item \ion{Ni}{2}$\lambda$1455 observed versus CMFGEN (\ref{sec:NiII}),
\item \ion{Ni}{2}$\lambda$1455  changes across periastrons 11 and 14 (\ref{sec:NiIIchange}),
\end{itemize}
\item Section \ref{sec:samephase} describes profiles  at similar binary phases: 
\begin{itemize}
\item profiles at nearly the same phase of three binary periods (\ref{sec:early}),
\item observed profiles of doublet members (\ref{sec:doublet}),
\item observed profiles compared to \cmfgen\ profiles (\ref{sec:wind}),
\item long term profile changes between periastron passages 11 and 14 (\ref{sec:longterm}),
\item profile changes across periastron passages 11 and 14 (\ref{sec:low}). 
\end {itemize}
\item Section \ref{sec:DIS} discusses these results.
\item Section \ref{sec:SUM} summarizes our conclusions and suggests future observations. 

\end{itemize}

\section {Setting the stage}\label{sec:stage}

\begin{figure*}
\includegraphics[width=18.cm]{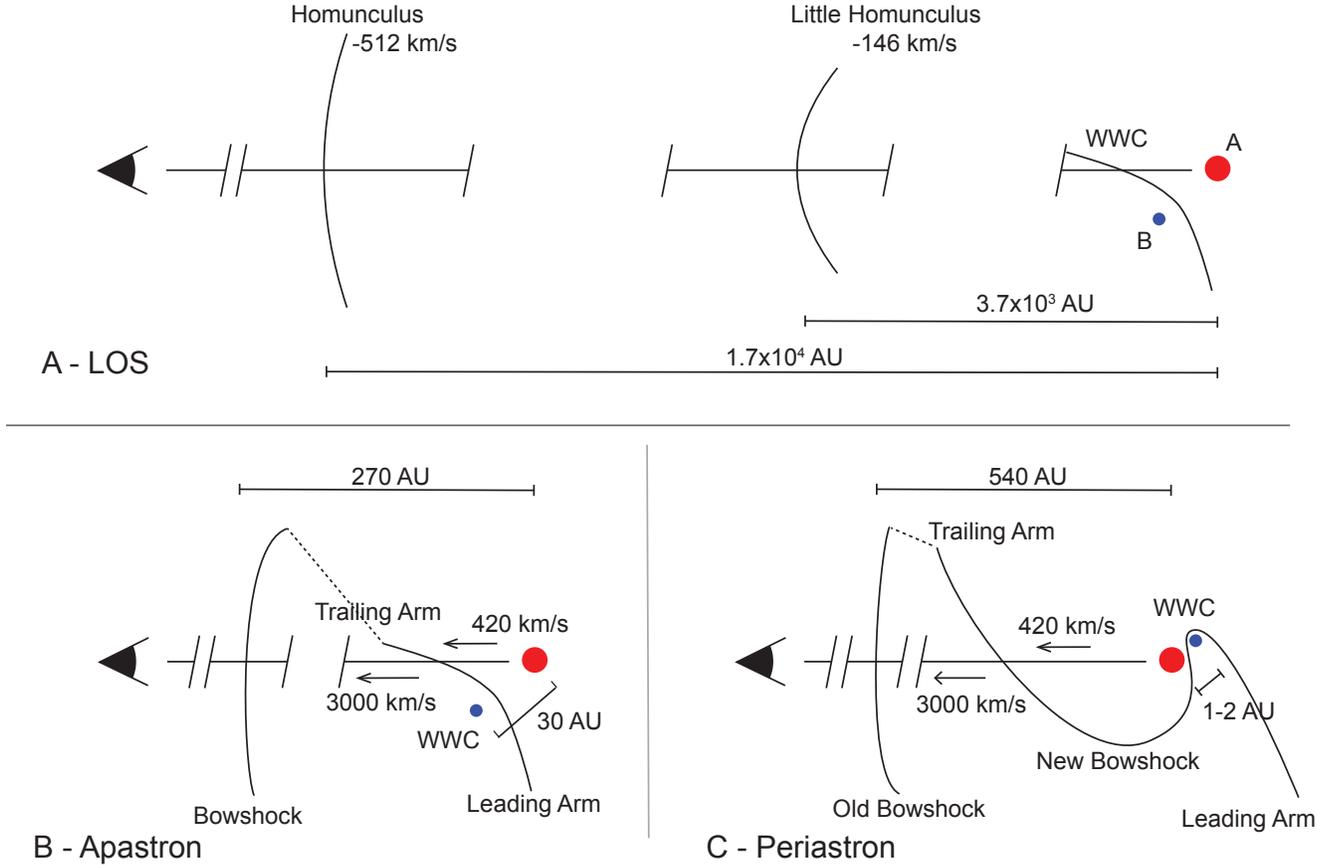}
\caption{{\bf Sketches defining the  structures in our LOS from \ec. A: Large-scale structure in LOS at apastron.} The binary companions are separated by $\approx$30 au with \ec-B on the near side of \ec-A.   Our LOS of \ec-A, the principle source of  {\bf radiation longward of Ly$\alpha$, 1216\AA,} passes through the highly shocked, wind-wind collision (WWC) zone close to the leading edge of the cavity carved by the wind of \ec-B out of the wind of \ec-A for most of the orbit.
In our LOS, two series of shells, the first characterized by the $-$146 \kms\ Little Homunculus,   3700 au from \ec,  and the second  characterized by the $-$512 \kms\ Homunculus,  170,000 au from \ec.  {\bf B: Enlargement of the interacting binary at apastron:} The stronger WWC zone, the X-ray source, is located between the two stars, separated by 30 au.  A secondary structure, the bowshock, exists at a distance of 270 au being the result of the secondary wind blowing a cavity out of the primary wind. {\bf C: Enlargement of the interacting binary near periastron:} The  WWC zone wraps around \ec-A as \ec-B passes to the far side. At a distance of 540 au, the bowshock continues to expand at $\approx$ 500 \kms\ in our LOS. For a brief interval at the beginning of the periastron passage, the trailing arm passes in front of \ec-A leading to a newly created cavity and bowshock in our LOS. {\bf Note that our LOS is at  49\degr to the binary orbital plane.}
}

\label{fig:concept}
\end{figure*}
\begin{figure*}
    \centering
    \includegraphics[width=17.cm]{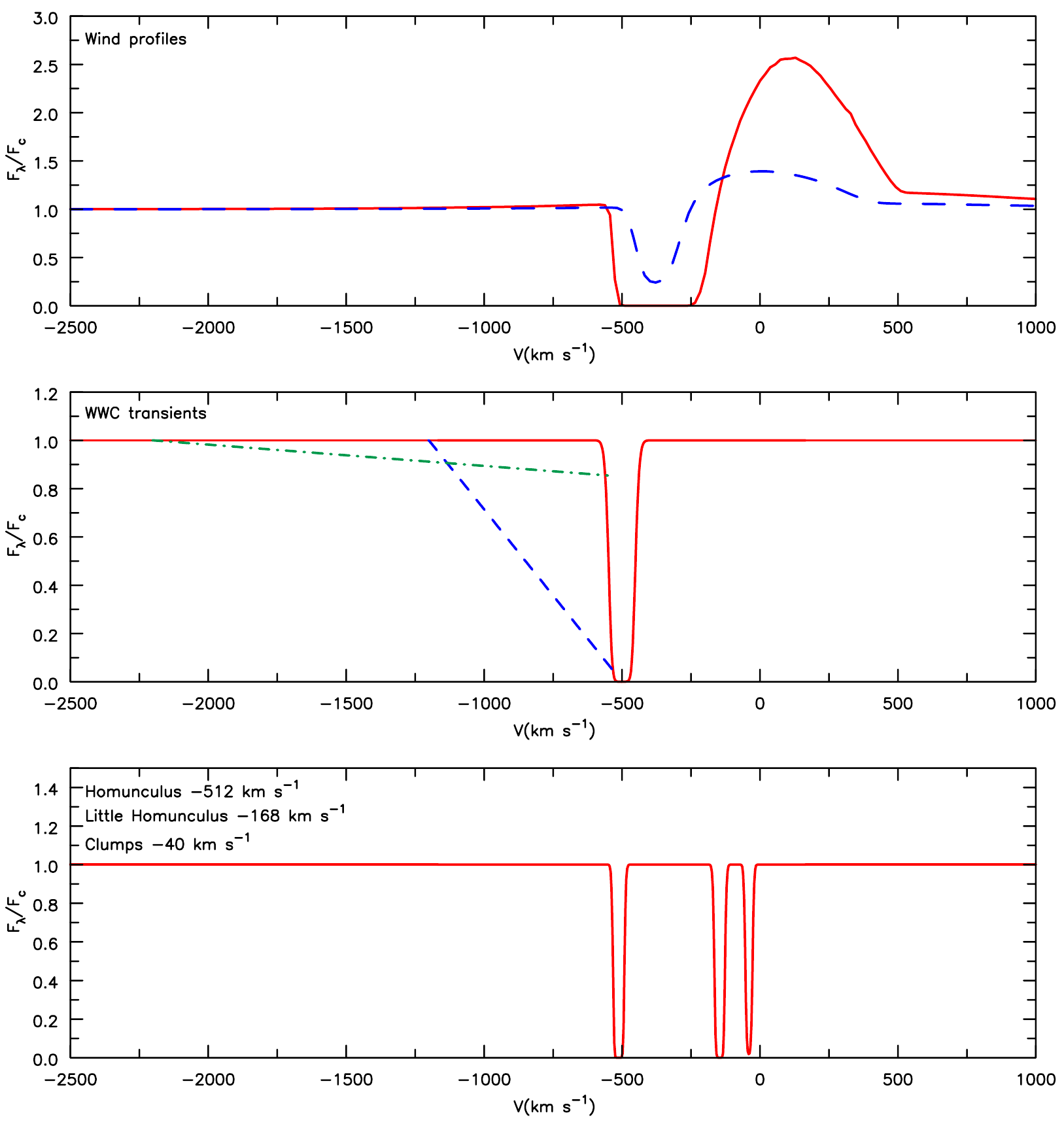}
    \caption{Absorption signatures of structures as sketched in Figure \ref{fig:concept} in our LOS. Top: \cmfgen\ model calculations predict saturated P~Cygni profiles ({\color {red}red}) and unsaturated
    profiles ({\color {blue}blue}) in transitions belonging to  lower ionization species such as \ion{Si}{2} and \ion{Ni}{2} but not in high ionization lines such as \ion{N}{5} or \ion{C}{4}. Middle: {\bf Representation of the variable WWC transient absorptions that} have been observed  in \ion{He}{1} $\lambda$10830 extending to $-$2200 \kms\ by \cite{Groh10} {\color {green} (green)} and in  FUV \ion{Si}{2} resonant lines exceeding $-$1200 \kms\ across the 2020 periastron passage by \cite{Gull21a} {\color {blue} (Blue)}.  Variable absorption from a portion of \ec-A's wind could also appear in high-ionization resonant lines (\ion{C}{4}, \ion{N}{5}) accelerated by the wind of \ec-B and highly ionized by the flux of \ec-B {\color {red} (red)}.  Bottom: Thirty-six narrow ($\approx$5.5 \kms\ FWHM) absorption velocities from shells were identified by \cite{Gull06}. {\bf The strongest absorptions, schematically shown, are at velocities $-$512 \kms\ (Homunculus), $-$146 \kms\ (Little Homunculus) and $-$40 \kms\  (Weigelt clumps located very near \ec).} With the ten-fold increase of FUV flux over the past two decades, many metals in these shells have increased from singly-ionized state to multiply-ionized states and the H$_2$ in the Homunculus has been  destroyed in our LOS.}
    \label{fig:absLOS}
\end{figure*}

The structures of \ec\ and its circumstellar
ejecta are far from simple. 
Three-dimensional hydrodynamical (3D hydro) models of the interacting winds, compared to  maps of spatially-resolved, visible-wavelength  forbidden emission lines, suggest that the binary orbital pole is closely aligned with the axis of symmetry of the bipolar Homunculus \citep{Gull09,Madura12, Gull16}. Our LOS intersects the orbital plane tilted 49\degr\ out of the sky plane.
The orientation of the stars around their orbits result in circumnebular ionization variations changes with binary orbital phase.  These changes are especially apparent across periastron passage due to the rapidly changing orientations of the two stars and consequent changes in the interacting winds. Longer-term changes in the circumnebular environment, and sporadic changes to the stars themselves, can result in non-phase-dependent variations in the spectrum as well.
As a result, changes in narrow line absorptions  associated with two large ejection events, are seen near periastron passage and over the long term. To aid our discussion, Figure \ref{fig:concept} shows a simple sketch that defines the structures near periastron and apastron along our LOS from the binary through its interacting winds  and circumstellar absorption shells.  Figure \ref{fig:absLOS} displays expected absorption signatures of these structures sketched in Figure \ref{fig:concept}.

The periastron event occurs with \ec-B buried deeply within the primary wind on the far side of \ec-A.  This leads to the observed low-ionization states of \ec-A and of the distant foreground shells within the  Homunculus every  periastron passage.  \ec-B is immersed in the primary wind for several weeks while the primary wind flows unimpeded in our direction.  \ec-B\ then emerges from the primary wind, and the secondary wind creates  a chaotic acceleration of primary wind in our direction \citep{Gull21a} and rebuilds the ionization cavity which is bounded by the swept-up primary wind. The very fast wind of \ec-B   creates a rapidly expanding, highly-ionized, low-density cavity and, in diluted form, accelerates and compresses the swept-up primary wind along the downstream walls of the bowshock.  Our LOS from the FUV source, \ec-A, views  the interior of the ionized cavity around \ec-B and the bowshock formed by the primary wind after each periastron passage.  Both the cavity and the bowshock continuously expand for most of the 5.54-year orbit, only to be replaced by a new cavity bounded by a new bowshock; in other words, a train of low density cavities bounded by bowshock walls accumulates and extends across our LOS. 

The core of \ec-A is the primary continuum source longward of \ion{H}{1} Ly$\alpha$ {\bf consistent with both observations and \cmfgen\ models} \citep{Hillier06}. However the hotter, less luminous \ec-B increasingly contributes flux at shorter wavelengths in the FUV and  dominates below \ion{H}{1} Ly$\alpha$ ($<$912\AA, the EUV) \citep{Iping05a}. As each periastron passage occurs,
the ionization and the excitation of the Weigelt clumps fall as they are cut off from \ec-B's
ionizing radiation by the inner wind of \ec-A \citep{Zethson01a,Zethson12,Hartman05,Verner05a}. A drop in ionization can also be seen in the circumstellar gas  along our LOS, producing the various absorption components \citep{Gull06}.

The continuum-emitting primary-wind core, with a diameter of 6.5 au in mid-IR Very Large Telescope Interferometer (VLTI) images taken near periastron, extends beyond the  1 to 2 au separation of the binary \citep{2021A&A...652A.140W}. Mid-IR measurement {\bf resolved the  \ion{H}{1} Br$\alpha$ which demonstrated that the  low-ionized portion of the primary wind extends to at least} 60 au \citep{Weigelt21}, well beyond the major axis of the binary orbit ($\approx$ 28 au).  

We identify absorption signatures of specific structures in our LOS as follow:
\begin{itemize}
    \item The  extended, cooler primary wind provides a P\,Cygni profile with  \Vinf$_{,A}$ = 420 \kms\ (Figure \ref{fig:absLOS}, Top). \cmfgen\ modeling suggests P\,Cygni profiles are present in the low-ionization resonant lines (Section \ref{sec:CMFGENmod}). Two example profiles are \ion{Ni}{2} and \ion{Si}{2}. \ion{C}{2},  \ion{Al}{2}, \ion{Al}{3} and possibly \ion{Si}{4}  also show evidence of P\,Cygni wind profiles (Figures  \ref{fig:NiIICMFGEN},\ref{fig:CMFGEN} and \ref{fig:CMFGENDoublet}).  
    \item The evolving bowshock provides absorption in our LOS (Figure \ref{fig:absLOS}, Middle): Following the 3D hydro modeling of \cite{Madura10,Madura12} the WWC exists in two spatially-separate regions: 
    \begin {itemize}
    \item The highly-shocked secondary gas between the binary stars gives rise to  X-rays \citep{2017ApJ...838...45C}. Across the high-ionization state, the leading arm of the shock is located in front of \ec-A. The secondary wind  inhibits the  flow of the primary wind in our direction but provides little absorption, while its EUV radiation leads to higher ionization of the primary wind. Little absorption is expected in the FUV during the high-ionization state from the shock boundary. 
    
    \item As the binary system approaches periastron passage, \ec-B plunges deep into the primary wind. The trailing arm of the WWC wraps around \ec-A leading to ionization and acceleration of the distorted primary wind. Both \cite{Groh10} and \cite{Richardson16} saw evidence of this trailing arm before periastron passages in cycles 12 and 13 in the form of high velocity absorption of \ion {He}{1} lines in the visible and the near red.  \cite{Groh10} also reported the presence of the trailing arm in FUV resonant lines as the binary approached periastron passage 11.
    In the FUV, multiple resonant lines show a transient absorption wing that is saturated at $-$500 \kms\ and  extends to $-$1000 to $-$1500 \kms (Figure \ref{fig:absLOS}, Middle).
    
    \item For several weeks or more when the secondary is near superior conjunction, the primary wind flows unimpeded  in our LOS. As \ec-B moves from behind \ec-A,  a new wind cavity becomes visible and the leading arm again passes in front of \ec-A. A bowshock then forms out of the downstream primary wind that is moderately accelerated by the secondary wind and more highly ionized by \ec-B's EUV radiation. The  apparent net result is  absorption from the accelerated wind of \ec-A over multiple ionization states extending from velocities of $-$400 to $-$600~\kms\ (Figure \ref{fig:absLOS}, Middle).
    \end{itemize}   
    Near periastron passage, the dense walls of the WWC are expected to contribute transient FUV absorptions extending to high velocities (as seen, for example, in the visible and near IR \ion{He}{1} lines by \citet{Groh10, Richardson16} and  in the FUV \ion{Si} {2} resonant lines by \cite{Gull21a}), along with absorptions that persist for much of the orbit.
    \item Narrow-line absorption at three dozen discrete velocities ranging from $-$122 to $-$1074 \kms\ were identified by \cite{Gull06} (Figure \ref{fig:absLOS}, Bottom). Some were at velocities near characteristic absorption velocities associated with the Little Homunculus ($-$146 \kms) or the Homunculus ($-$512 \kms), but some were at even higher blueshifted velocities. 
    \end{itemize}

The dissipation of the occulter, located within the Little Homunculus (Figure \ref{fig:concept}),   causes
systematic, long-term changes in the observed stellar spectrum. Because it preferentially  blocked the central stars and the WWC region more so than the extended primary wind, the dissipation of the occulter has lead to changes in the spectrum, even if the stellar fluxes are constant.
In the visible, the dissipation of the occulter
has led to a decline in the EWs of many \ion{H}{1}, \ion{Fe}{2} and [\ion{Fe}{2}] emission lines,  since the observed continuum has increased. 
It has also influenced the strength of the  P~Cygni absorption profiles associated with \ion{Si}{2} and \ion{N}{2} \citep{Damineli21}, (Damineli et al. submitted).

 \section{The Observations} \label{sec:OBS}
 
 \begin{center}
 \begin{table}
\tablenum{1}
\caption{Log of \hst/\stis\ Observations$^a$\label{tab:obs}}
\begin{tabular}{llll}
\hline\hline
 Date      &Phase$^{b}$ & Aperture$^{c}$       & Gratings\\
\hline
2002-01-20 & 10.736 & 0\farcs3$\times$0\farcs2 & E140M, E230M\\
2002-07-04 & 10.818 & 0\farcs3$\times$0\farcs2 & E140M, E230M\\
2003-02-13 & 10.928 & 0\farcs3$\times$0\farcs2 & E140M, E230M\\
2003-05-26 & 10.979 & 0\farcs3$\times$0\farcs2 & E140M, E230M\\
2003-07-01 & 10.998 & 0\farcs3$\times$0\farcs2 & E140M, E230M\\
2003-07-29 & 11.010 & 0\farcs2$\times$0\farcs2 & E140M, E230M\\
2003-09-21 & 11.037 & 0\farcs3$\times$0\farcs2 & E140M, E230M\\
2004-03-06 & 11.119 & 0\farcs3$\times$0\farcs2 & E140M, E230M\\
2015-09-02 & 13.194 & 0\farcs2$\times$0\farcs2&E140M, E230M\\
2018-04-21 & 13.670 & 0\farcs2$\times$0\farcs2&E140M, E230M\\
2019-06-10 & 13.886 & 0\farcs2$\times$0\farcs2&E140H\\
2019-08-25 & 13.913 & 0\farcs2$\times$0\farcs2&E140H\\
2019-12-28 & 13.975 & 0\farcs2$\times$0\farcs2&E140H\\
2020-02-08 & 13.995 & 0\farcs2$\times$0\farcs2&E140H\\
2020-03-09 & 14.010 & 0\farcs2$\times$0\farcs2&E140H\\
2020-03-11 & 14.011 & 0\farcs2$\times$0\farcs2&E140M, E230M\\
2020-04-01 & 14.022 & 0\farcs2$\times$0\farcs2&E140H\\
2021-01-26 & 14.170 & 0\farcs2$\times$0\farcs2&E140M, E230M\\
\hline
\end{tabular}
\\
$^{a}$ Additional \hst/STIS FUV spectra of \ec\ were recorded, but we only list those at similar phases along with additional spectra that are key to tracking changes in  resonant-line profiles relevant to this discussion.\\
$^{b}$ Phase refers to the binary orbital phase based on both disappearance of He I emission and X-ray drop with periastron passage, numbered 13 occurring on JD 2456874.4 $\pm$ 1.3 days, and orbital period of 2022.7 $\pm$ 0.3 days  \citep{Teodoro16}. The periastron number refers to the convention  based upon  detected periastron passages via visible spectroscopy beginning February 1948. While there is general agreement on the period, observations and models do not concur on the actual time of the periastron event, which is thought to be several  days earlier than this reference date. \\
$^{c}$ Spectra using the unsupported aperture 0\farcs3$\times$0\farcs2 were extracted with STIS-team software for a stellar source excluding extended wind background. For details, see {\cite{Gull21a}}.
\end{table}
\end{center}

A series of  \hst/STIS  high-dispersion FUV spectra was recorded in coordination with {\it Chandra} observations following the spectroscopic X-ray changes across \ec's periastron event in 2020.2 \citep{espinoza21b}; (Espinoza et al., accepted). The dates of observations and relevant binary orbital phases are listed in Table \ref{tab:obs}.
The \hst/STIS visits were planned close to the scheduled {\it CHANDRA} visits (1) to establish a baseline of spectral changes as the binary approached periastron, and (2) to sample changes across periastron, following the drop to  low ionization-state of the interacting winds and the Homunculus shells when the hot secondary star is embedded deep within the primary wind and and the EUV flux of \ec-B  is blocked. 

A previous series of STIS FUV spectra had been recorded through multiple programs. Many of these spectra{\bf, as they were recorded as similar binary orbital phases, lent themselves to direct comparisons (Table \ref{tab:obs})}..

The observed FUV flux of \ec\ has increased tenfold since 2000 when initial observations were obtained with  \hst/\stis.  
Because of the increasing brightness of \ec, the 
medium-resolution echelle modes (E140M and E230M)  previously  used  to monitor the spectral changes from 1150 to 2350\AA\ could not be used for these new observations since the predicted global count rate 
approached  the safety limits of the Multi-Anode-Multi-Array (MAMA) detectors. We therefore switched to the higher-dispersion echelle mode, E140H, which unfortunately limited observations acquired in a single \hst\ orbit visit to  three cross-dispersed grating settings. We  chose  to monitor the most important resonant lines located between 1150 and 1680\AA\, with three grating settings: 1234, 1416 and 1598\AA.  This meant we were unable to  monitor other resonant lines of interest,  such as the \ion{Al}{3} $\lambda\lambda$1855,1863 doublet in the mid-ultraviolet spectral region covered by the E230M grating setting. 

Where applicable to this study, we included other \hst/STIS FUV spectra of \ec\ from  the public {\it MAST} archive. We also included archived E230M spectra  to examine the profiles of \ion{A}{3} $\lambda\lambda$1855,1863. Additional details on the stellar spectral extractions, specifically from the observations that utilized the 0\farcs3$\times$0\farcs2 aperture are described in \cite{Gull21a}.

\section{Resonant lines and what they tell us about structures in the LOS}
{\label{sec:resonant}}
 
\subsection{Wind lines of Interest}\label{sec:windlines}

\begin{table*}
\tablenum{2}
\centering
\caption{Resonant lines and element ionization potentials$^{a}$\label{tab:lines}}
\begin{tabular}{lrrrll}
\hline
 Spectrum  & log $gf$ & \multicolumn{2}{c}{Excitation energy} & Wave- & Velocity$^b$\\
 & & Lower & Upper & length & range\\
 &   & eV & eV & \AA & \kms\\
 \hline
\hline
\ion{N}{5}&$-$0.505&0.000 & 10.008 & 1238.821&$\approx$2000\\
\ion{N}{5}&$-$0.807&0.000 & 9.976 & 1242.804& 915\\
\hline
\ion{C}{4}& $-$0.419&0.000 & 8.008 & 1548.187&$\approx$3000\\
\ion{C}{4}&$-$0.720&0.000 & 7.995 &1550.774&498\\
\hline
\ion{Si}{4}&0.011&0.000 & 8.896 & 1393.755&$\approx$4000\\
\ion{Si}{4}&$-$0.292&0.000 & 8.896 &1402.770&1939 \\
\hline
\ion{C}{2}&$-$0.589&0.000 & 9.290 & 1334.532&$\approx$4000\\
\ion{C}{2}&$-$1.293&0.008 & 9.290 & 1335.663\\
\ion{C}{2}&$-$0.335&0.008 & 9.290 & 1335.708& 284\\
\hline
\ion{Si}{2}&$-$0.575&0.000 & 8.121 & 1526.707&$\approx$4000\\
\ion{Si}{2}&$-$0.274&0.036 & 8.121 & 1533.431&1320\\
\hline
\ion{Al}{2}&$-$0.248&0.000 & 7.421 & 1670.842&$\approx$4000\\
\hline
\ion{Al}{3}&$+$0.050 &0.000 & 6.685 & 1854.716&$\approx$4000\\
\ion{Al}{3}& $-$0.253&0.000 & 6.656 & 1862.789 & 1305\\
\hline
\ion{Ni}{2}$^{c}$&$-$0.447&0.000 & 8.450 &1454.842&$\approx$3000\\
\ion{Ni}{2}&-1.176&0.000&8.450&1467.259\\
\hline
\end{tabular}
\\
\begin{center}
\begin{tabular}{lrrrrrr}
\hline \hline
Element & \multicolumn{6}{c}{Ion charge$^a$} \\
 & 0 & +1 & +2 & +3 & +4 & +5 \\
&eV&eV&eV&eV&eV & eV \\
\hline
Ne & 21.5 & 41.0 & 63.4 &  97.2 & 126 & 158\\
Ar & 15.8  & 27.6 & 40.7 & 59.6 & 75 & 91 \\
N & 14.5 & 29.6 & 47.4 &  77.5  &  98 & 552 \\
C  & 11.3 & 24.4 & 47.9 &  64.5 & 392 & 490 \\
Si & 8.2  & 16.4 & 33.5 &  45.1 & 167 & 205 \\
Fe & 7.9  & 16.2 & 30.7 & 54.9 & 75 & 99 \\
Ni & 7.6 & 18.2 & 35.2 &  59.4 &  76 & 108 \\
Al & 6.0  & 18.8 & 28.5 & 119.3 & 154 & 190 \\
\hline
\end{tabular}
\caption*{
$^{a}$Lines and ionization potentials from \href{https://physics.nist.gov/PhysRefData/ASD/}{NIST}; 
$^{b}$Upper value is an estimate of the quasi-clear velocity interval before strong absorption occurs. Lower value is the effective velocity interval between doublet members due to spectral separation. (The \ion{Al}{2} line does not have a doublet companion). 
$^{c}$ From the  \href{https://lweb.cfa.harvard.edu/amp/ampdata/kurucz23/sekur.html}{Atomic spectral line database} (Kurucz)}
\end{center}
\end{table*}

The FUV interval includes an abundance of resonant lines from neutral as well as singly- and multiply-ionized elements. 
Indeed much of the challenge is sorting out the contributions from various line transitions and identifying the emitting and absorbing structures in the LOS. The lines discussed  are given in Table \ref{tab:lines} along with the ionization potentials (IP) for selected atomic/ionic species. For most of the transitions,  the lowest level of the transition is either the ground term, or a low-lying level which is easily excited. Thus the relevant IP for observing a particular ionization stage is that belonging to the previous ionization stage. For example, to get the \ion{C}{4} resonance transitions we generally need photons (or electrons) with energies greater than 64.5 eV (i.e., the IP of C$^{+3}$). In the inner wind, where densities are high and the radiation not strongly diluted, slightly lower photon energies may suffice since in some cases the ionization can be maintained via ionizations from excited states \citep{Johansson04}. In some cases it is
also possible that a species will give rise to a strong UV profile, even though it is never the dominant ionization stage.

\ec-A's  prominent wind lines are from singly-ionized carbon, silicon, aluminum, iron, and nickel, plus doubly-ionized aluminum. In the primary wind the ions giving rise to these lines can all be produced via photoionization from the primary. However, for the outer ejecta 
(e.g., Weigelt clumps, Little Homunculus) an additional source of radiation is needed. 
This is particularly true for doubly-ionized species and any species with an IP close to that of hydrogen (e.g. neutral carbon).  This  occurs because hydrogen recombines  in the outer wind of the primary without the radiation from the companion star. Narrow emission lines of [\ion{Fe}{3}], [\ion{Ne}{3}], [\ion{S}{3}], and [\ion{Ar}{3}] from the Weigelt clumps disappear at periastron when the secondary star goes behind the primary \citep{Johansson04a,Damineli08b, Gull09}. Absorption lines of \ion{C}{1} and \ion{N}{1} are also seen in the spectra, especially below 1350\AA, but that portion of the spectrum is so crowded that only very limited intervals in velocity can be studied for some specific resonant lines.

In addition to absorption  lines from {\bf singly-ionized and neutral} species,  resonance doublets of higher-ionization species, like \ion{Si}{4}, \ion{C}{4}, and \ion{N}{5} are also seen (Section \ref{sec:CMFGENmod}). The high-ionization \ion{Si}{4} and \ion{C}{4} P~Cygni profiles, and the weaker absorption associated with \ion{N}{5}, cannot be produced by photoionization from the primary star. These lines must either be associated with the secondary wind, the wind-wind shocked region or other outflowing gas bathed in the UV radiation field of the secondary and/or gas illuminated by X-rays arising from the wind-wind shock. 

The resonant absorption lines trace the ionization of different structures in our LOS. For example, the \ion{C}{2} $\lambda\lambda$1335,1336 absorptions occur in gas 
in the ionization range from 11.3 to 24.4 eV, and likewise the \ion{C}{4} $\lambda\lambda$1548,1551 absorptions trace gas at an ionization state above 64.5 eV. \ion{N}{5} $\lambda\lambda$1239,1242 absorptions originate in gas ionized above 98 eV. Hence we have a means of estimating the range of ionization for structures in the LOS ranging from 6.0~eV (\ion{Al}{2} 6.0 to 18.8 eV) to $>98$~eV (\ion{N}{5} $>$98 eV), as shown 
in Table \ref{tab:lines}. 

Most of the lines studied in this paper have nearby companion lines due to transitions to a second upper level or from an energy level close to the ground state. \ion{N}{5}, \ion{C}{4}, \ion{Si}{4} and \ion{Al}{3} have  resonances to two different upper levels. \ion{Si}{2} and \ion{C}{2} have a low-lying energy level which can be thermally populated. Resonant lines are  seen in cold, interstellar clouds\footnote{An example is a foreground cloud at $+$87 \kms\ that is seen in \ion{Si}{2} $\lambda$1527 and \ion{Al}{2} $\lambda$1671.} but the companion line is absent. However, the companion line is present in the  hotter wind structures and multiple discrete shells of gas within the Homunculus. Among the resonant lines with well-defined profiles, only the \ion{Al}{2} $\lambda$1671 line lacks a companion.

Doublet line separations range in velocity from 280 \kms\ 
(\ion{C}{2})\footnote{The \ion{C}{2} $\lambda$1335 resonant line is a blend of three \ion{C}{2} lines (Table \ref{tab:lines}): 1334.532\AA, 1335.733\AA\ and a much weaker line, 1335.663\AA. The latter two components lie within 14\,\kms\ of each other, so they are treated as a single line in the present paper.} to 1320 \kms\ (\ion{Si}{2}). Limited strong absorptions from other lines probe relatively clear velocity intervals up to 4000 \kms\ shortward of the blue doublet line. While the \ion{C}{2} doublet separation is insufficient to separate individual line absorption across a 600 \kms\ range, the \ion{C}{2} doublet can be used to track  variable high-velocity absorptions extending as negative as $-$1900 \kms\ \citep{Groh10}. We use these resonant doublets to help understand the influence of line blends, to search for absorptions up to $-$3000~\kms\ (the terminal velocity of the \ec-B's wind) and to study the 0 to $-$600 \kms\ interval which is strongly affected by broad absorptions  ($\Delta~V\approx$ hundreds of \kms\ in width) from the primary wind, which are 
much broader in velocity  then the circumnebular absorptions \citep[$\Delta~V\approx$ 5 to 10~\kms, ][]{Gull06}, and which often have associated emission as well.
\subsection {Description of the \cmfgen\ model}\label{sec:CMFGENmod}

The adopted model is the same used to interpret Very Large Telescope Interferometer (VLTI)  observations by  \cite{Weigelt21}.  The parameters are similar to, but not identical to those found by \citep{Hillier01,Groh12}. A slightly larger radius was adopted than in earlier studies to weaken \ion{He}{1} lines -- these lines are now  believed to have a substantial contribution  from the bow shock \citep[e.g.,][]{Nielsen07b}. Even larger radii can be used to explain the spectroscopic observations, but these are probably ruled out by the high eccentricity of the binary orbit. With $e=0.9$ and $a=15\,$AU, the periastron distance of 1.5\,AU is already uncomfortably close to the radius of the  primary star (0.56\,AU).
 
In our modeling we assumed $d=$2.3\,kpc \citep{Allen91}, which is very similar to the most recent Gaia EDR3 determinations of distance to  the star clusters Trumpler 14, 15 and 16 \citep{Shull21}.   In practice revised stellar parameters for a new distance can be obtained through a simple scaling ($L\propto d^2$, $R\propto d^2$, $\Mdot \propto R^{3/2}$). A detailed discussion of the complex issues surrounding the modeling of the primary star,
and the resulting uncertainties, will be provided in another paper (Hillier at al, in preparation). 
 
The stellar parameters and abundances model are listed in Tables \ref{tab_params} and \ref{tab_abund}. While the model provides a reasonable match to the optical spectrum, a comparison of the predicted UV spectrum to the observed spectrum shows a very poor match. This is  primarily due
to the influence of the companion star which carves out a cavity in the wind of the primary star \citep{Groh12}. During most of the orbit it is believed that we are looking at \ec-A through this cavity which has different ionization structure and density than the primary wind. The secondary star can also modify the ionization structure of the primary wind by its UV radiation.  The main consequence of the above two effects is that the absorption spectrum, particularly due to \ion{Fe}{2}, is much stronger in the model than in the observations. For UV data taken by HST in late 1980's and 1990s there was an additional effect. The coronagraph (herein referred to as the occulter)
was blocking much of the stellar UV light from reaching us, and what we observed was mostly the outer halo around the binary system \citep{Hillier03}. Fortunately the coronagraph has much less influence today.
\begin{table}
\tablenum{3}
\caption{Star and wind parameters used in model}
\begin{center}
\begin{tabular}{ll} 
\hline\hline
Parameter &     Value\\
\hline
L       & $4.0\times 10^6$\,\Lsun \\
\Mdot   & $8.0 \times 10^{-4}$\,\Msunyr \\
$R$\tauv{10} & $1.28 \times 10^2$\,\Rsun \\
$R$\tauv{2/3} & $7.42 \times 10^2$\,\Rsun  \\
T$_{eff}$\tauv{10} & $2.28 \times 10^3$\,K \\
T$_{eff}$\tauv{2/3} & $9.47 \times 10^2$\,K \\
$V$\tauv{10} & $ 5.4 \times 10^1$\,\kms \\
$V$\tauv{2/3} & $ 3.57 \times 10^2$\,\kms  \\
\Vinf               & $4.20 \times 10^2$\,\kms \\
\hline
\end{tabular}
\end{center}
\label{tab_params}
\end{table}

\begin{table}
\tablenum{4}
\caption{Model abundances}
\begin{center}
\begin{tabular}{lcc} 
\hline\hline
Species & Rel. Num.   & Mass  \\
              & Fraction & Fraction \\
\hline
H       & 5.0    &  0.547  \\
He     & 1.0    &  0.438  \\
C       & $1.7\times 10^{-4}$  &  $2.23 \times10^{-4}$   \\
N       & $6.8 \times  10^{-3}$  & $1.04\times 10^{-2}$ \\
O      &  $1.0 \times  10^{-4}$   &  $1.75 \times 10{-4}$  \\
Al      &  $2.2 \times 10^{-5}$  &   $6.36 \times 10^{-4}$  \\
Si      &  $2.9 \times 10^{-4}$  &   $8.77 \times 10{-4}$ \\
Fe     &  $2.5 \times 10^{-4}$   &   $1.53 \times 10^{-3}$  \\
Ni      &  $1.1 \times 10^{-5}$  &     $7.32 \times 10 ^{-5}$ \\
\hline
\end{tabular}
\label{tab_abund}
\end{center}
\end{table}
For this paper we took into account the cavity carved into the wind of \ec-A's wind using the analytical model of \cite{Canto96}. To compute the location of the cavity,
and its shape, we assumed a mass-loss rate of $1.0\times\,10^{-5}\,\Msunyr$ and a terminal velocity of 3000\,\kms\ for the secondary  \citep{Pittard02}. In the cavity we simply reduced the opacities and emissivities of the primary wind by a factor of 100. We made no attempt to include the secondary star and its wind, or the bow shock.

We then took the pre-existing model, and used a 2D radiation transfer code to compute the observed spectrum as a function of inclination, where (for our axisymmetric system) an inclination of zero degrees means that we are looking along the binary plane of the system with star B in the foreground. Our approach is very similar to that of \citep{Groh12}.
Crudely, the resulting spectra fall into 2 classes -- those
where we look through the cavity, and those where we look only through the primary wind. It is the former that gives the best agreement with observations.

The aim of our work was not to produce a fit to the UV spectrum -  rather it was designed to give a theoretical  spectrum that could be used to help interpret the complex UV 
spectrum that we observe. Given the simplicity of the approach, and the neglect of the influence of the bow shock and the  secondary UV radiation,  it provides a surprisingly good match to the UV spectrum. Its main deficiencies are that it does not explain the \ion{S}{4} and \ion{C}{4}
P~Cygni profiles, nor absorption velocities that exceed  $\sim-$500\,\kms.

At periastron, the secondary star moves behind the primary  and does not carve a cavity in the wind of the primary. Thus one might naively expect that the model spectrum would closely resemble that at periastron. This is true for many optical lines, but not for the UV and H$\alpha$. The UV and H$\alpha$ formation region is large \citep{Hillier01,Hillier03} and the flow time scales need  to be allowed for. For example, at \Vinf\ $=$ 500\,\kms, 30 AU (or $\sim 100\,R_*$) corresponds to a flow time scale of approximately 100 days.


\subsection{The \ion{Ni}{2} $\lambda$1455 line}{\label{sec:NiII}}

\begin{figure}
\centering
\includegraphics[width=8.5cm]{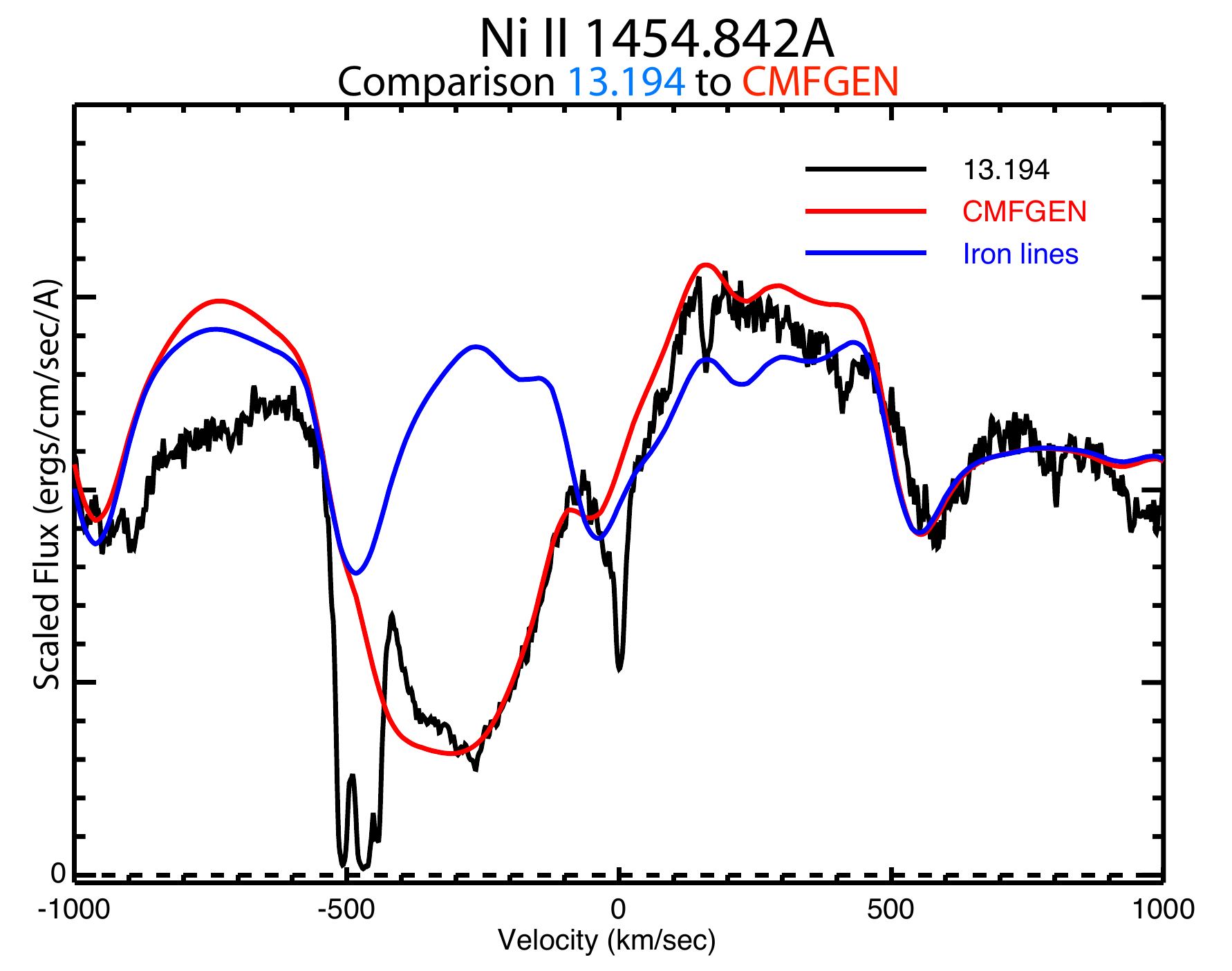}
\caption{  \ion{Ni}{2} $\lambda$1455 velocity plot from {\bf $\phi =$ 13.194} compared to {\red a \cmfgen-generated model of \ec-A}  and {\color {blue} the same model but showing only line contributions by iron.} The \cmfgen-generated model reproduces the UV spectrum of \ec\ remarkably well. Most of the UV portion of the \ec\ spectrum is heavily modulated by low-ionization iron lines as notable here. The \cmfgen\ velocity profile, generated for \Vinf\ $=$ 420 \kms\ \citep{Groh12}, closely follows the \ion{Ni}{2}$\lambda$1455 velocity profile quite well. Exceptions are caused by two velocity systems originating from ejecta: narrow \ion{Ni}{2} $\lambda$1455 centered at $+$4 \kms\ blended with an iron absorption to the blue and a complex of absorptions above and below the $-$513 \kms\ velocity of the Homunculus \citep{Gull06}.}
\label{fig:NiIICMFGEN}
\end{figure}

In this section we examine changes in \ion{Ni}{2} $\lambda$1455, an example of a broad P\,Cygni low-ionization resonant line originating in the primary wind.
We discuss  higher velocity, broad absorptions in both low- and high-ionization resonant lines that originate from the wind-wind interactions and the secondary wind at velocities of $-$600 to  $-$2600 \kms. 
Figure \ref{fig:NiIICMFGEN} shows the \ion{Ni}{2} $\lambda$1455 line at $\phi =$ 13.194 along with a model using CMFGEN (see Section \ref{sec:CMFGENmod}). 
Although it is one of the very few relatively isolated wind lines in the FUV spectrum of \ec, \cmfgen\ modeling  indicates that iron lines from the primary's wind still influence the profile, and absorptions from  intervening ejecta and interstellar gas also contribute. The {\color{red} red tracing} in Figure \ref{fig:NiIICMFGEN} is the  full \cmfgen\ spectrum and the {\color{blue} blue tracing} isolates the iron contributions which dominate the UV spectral region.

From Table \ref{tab:lines}, the first and second ionization energies of nickel are 7.64\,eV  and  18.17\,eV, below and above the ionization potential of hydrogen (13.6\,eV). 
This resonant line provides a means of tracking the lower ionization boundary of the primary wind and the ionized ejecta within the Homunculus and Little Homunculus.
The line shows broad, unsaturated absorptions  extending from 0 to $-$600 \kms\  with a broad emission component that peaks around $+$200 \kms, extending redward to $+$500 \kms. The large offset of the emission peak velocity is most likely due to blending with low velocity absorption. 

\begin{figure}
\centering
\includegraphics[width=8.5cm]{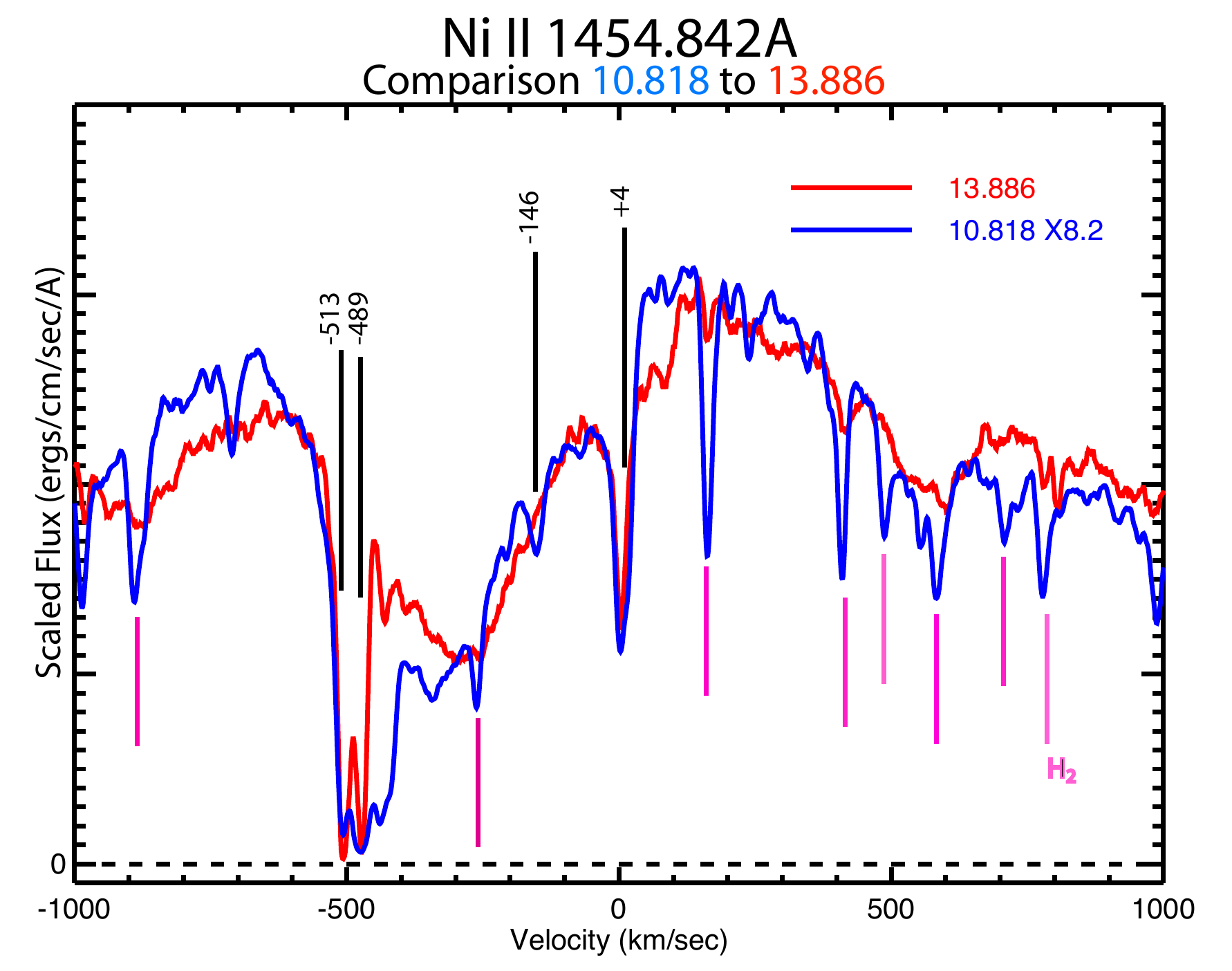}
\caption{\ion{Ni}{2} $\lambda$1455 velocity plots at similar binary phases of {\color{blue} $\phi =$10.818} and  {\color{red} $\phi =$13.886}. The flux of the $\phi =$ 10.818 spectrum has been scaled to the flux at $\phi =$ 13.886  at $\lambda =$ 1483\AA, a relatively line-free portion of the spectrum. Between the two observations, 16.5 years apart, the apparent flux has 
increased by a factor of 8.2. The increase in FUV flux  led to the disappearance of 1) H$_2$  absorptions at $-$512 \kms\ (Homunculus) and 2)  \ion{Ni}{2}\,$\lambda$1455 absorptions  at $-$146\, \kms\ (Little Homunculus) and from $-$250 to $-$475 \kms\ from the multiple shells within the Homunculus and Little Homunculus. Strong  absorption remained at $-$486 and $-$513 \kms, the stronger shells at the boundary of the Homunculus. The shape of the P\,Cygni wind profile has not changed significantly.}
\label{fig:NiII10to13}
\end{figure}

Figure \ref{fig:NiII10to13} compares two profiles of \ion{Ni}{2} $\lambda$1455. Narrow absorptions are seen at $-$513 \kms\ (arising in the expanding Homunculus) and at $-$146 \kms\  \citep[arising from the Little Homunculus,][]{Gull05}. The  P\,Cygni profile peaks at $+$200 \kms\ with an unsaturated absorption reaching minimum near $-$300 \kms\ in spectra at phases $\phi =$ {\color{blue}10.818} and {\color{red}13.886} (i.e. at nearly the same orbital phase 16.6\,years apart). Four narrow \ion{Ni}{2} $\lambda$1455 absorption systems are in our LOS: one interstellar, $+$4 \kms and three absorptions from ejecta shells arising from the  Little Homunculus (at $-$146 \kms), the Homunculus( at $-$513 \kms), and near the Homunculus, $-$489 \kms\ \citep{Nielsen08a}. 

While the $-$146 \kms\  component disappeared by $\phi =$ 13.886, some weaker, singly-ionized metal  lines briefly returned during periastron passage 14. The optical resonance line, Na D, is discussed by Pickett et al. (in prep), who found that the $-146$~\kms\ component weakened after 2009 (periastron passage 11) and disappeared in 2015 and 2016 (following periastron passage 12), while a $-168$~\kms\ component appeared briefly during the 2020 periastron passage.

Eight narrow lines of H$_2$ at $-$513 \kms\ are present in the {\color{blue}$\phi =$10.818} velocity plot but are not seen at {\color{red} $\phi =$ 13.886}. These absorptions are members of the ultraviolet Lyman  bands. Ten percent of the band transitions  lead to molecular dissociation \citep{Dalgarno70, Jura74}. Model calculations by \cite{Werner69}, scaled to the changing radiation field within the Homunculus, suggest that the dissociation lifetime of N$_{H2}$ dropped from a century to a decade between 2002 and 2019, likely the direct result of the tenfold increase in FUV radiation.  Across  periastron passage 11, the H$_2$ absorptions briefly disappeared, but returned by phase 11.119 \citep{Nielsen05a}. The transient disappearance of H$_2$ absorptions is due to absorption of the FUV flux (necessary to  populate upper  H$_2$ levels) by the intervening primary wind across the periastron passage.

The disappearance of lines of singly-ionized elements across the high ionization state
occurred because of the increase in ionization (e.g., Ni$^+$ to  Ni$^{++}$, Ti$^+$ to  Ti$^{++}$, etc.) due to the  increased photoionizing radiation in our LOS.
The increase in  photoionizing radiation field has produced many changes in the narrow-line absorption spectra since periastron 11 in 2003.5.
Details of the changes in the multiple shells are a topic for another paper (Nielsen et al., in prep). Despite the dramatic increase in the overall UV flux level, the  P\,Cygni profiles 
have not substantially changed (after allowing for the high-velocity components from the several shells within the Homunculus along the LOS).  

\subsection{Comparison of Ni~II~1455\AA\ changes between periastron passages 11 and 14}{\label{sec:NiIIchange}}

 \begin{figure*}
\includegraphics[width=18cm]{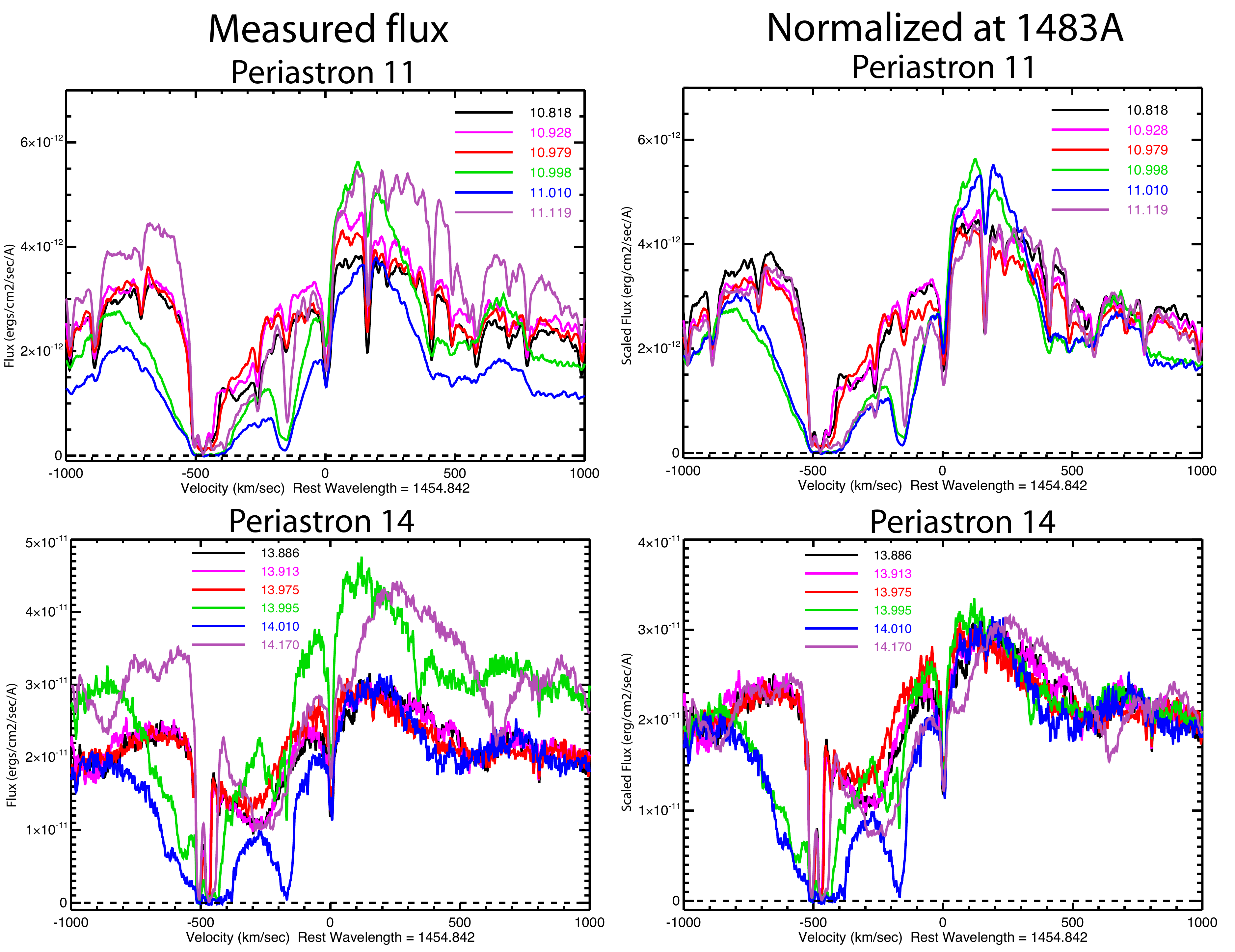}
\caption{{\bf Velocity profiles of \ion{Ni}{2} $\lambda$1455 changing across periastron passages 11 (top row) and 14 (bottom row).} Profiles plotted in the left column change in amplitude due to the changing continuum levels. Profiles in the right column have been normalized to  the continuum level near 1483\AA\ (1475 to 1490 \AA).
The normalized profiles (right column) show very little change leading up to periastron (10.818, 10.928 10.979; 13.886, 13.913;  13.975) but change dramatically across periastron passage (10.998, 11.010; 13.995 14.010) where there is a
general enhancement in absorption, with extended absorption from  $-$500 to $-$800 \kms\ now present.
Profiles are returning  to the pre-periastron values by phases 11.119 and 14.170.}
\label{fig:NiIItrans}
\end{figure*}

In this section we study the behaviour of \ion{Ni}{2} $\lambda$1455 across periastron passages 11 and 14. Due to variations in continuum flux both within an orbital cycle and over longer timescales,  Figure~\ref{fig:NiIItrans} shows the \ion{Ni}{2} $\lambda$1455 profiles in two ways:  In the left column the profiles are displayed 
in  absolute flux,  while in the right column the spectra are 
 scaled by the measured continuum from  1475 to 1490\,\AA\ , which shows minimal spectral changes over an orbital period \citep{Groh10}. 

Direct  comparison of the profiles at each phase is complicated by the
changes in flux between cycles 10  and 13 and the changes in the strength of the absorptions arising in the Homunculus and Little Homunculus. However, a comparison
of flux-scaled profiles (Figure~\ref{fig:NiIItrans}, right column) reveals that, despite the nearly tenfold change in flux, the overall shapes of the profiles near the high-ionization state are similar, as are profiles in the low ionization state (near $\phi=11.01$ and $\phi=14.01$).

The scaled spectra provide  much clearer views of the changes in the resonant line profile across periastron passages 11 and 14. By $\phi =$ 10.998 (13.995), strong,  broad absorption  developed  blue-shifted beyond  $-$600 \kms\  extending to $-$800 \kms. This broad absorption disappeared in the recovery stages  in both cycles ($\phi =$ 11.119 \& 14.022).
The  continuum-normalized profiles are sensitive to changes in ionization of the colliding wind bowshock, so the similarity in profile shape between cycle 11 and cycle 14 suggests little change in ionization state of the wind-wind collision. 

Across periastron 11 (Figure \ref{fig:NiIItrans} top left), the continuum UV flux {\bf declined} by $\phi =$ 11.010 but  brightened by $\phi =$ 11.119, to  higher levels than those measured at $\phi=$ 10.979 \citep{Gull21a}. In contrast, across periastron 14 (Figure \ref{fig:NiIItrans}, bottom left), the continuum {\bf increased} by $\phi =$ 13.995, faded somewhat by $\phi =$ 14.010 and recovered to pre-periastron flux level by $\phi =$ 14.022). 

Some orbit-related changes in the UV in each cycle are linked to the ``borehole'' effect \citep{Madura12B, Gull21a}, a cavity deep in the inner wind of  \ec-A near periastron passage, produced when \ec-B enters the inner wind of \ec-A, which exposes deeper, hotter layers of \ec-A that leads to an increase in ionizing continuum for a brief interval.  
The borehole produces a strengthening of the red-shifted broad emission across both periastron passages in the velocity range from $+$50 to $+$300 \kms\ when the extended wind moves into the borehole cavity. Eventually the primary wind then fills in the borehole
and the red-shifted emission drops towards equilibrium.
The similarity of the borehole effect in the two cycles suggests that the intrinsic flux from the inner regions of \ec-A has not changed appreciably in the 16 years  separating the two orbital cycles.

Ionization of Ni to Ni$^+$ (IP $=$ 7.6\,eV) is easily provided by UV radiation from \ec-A. However ionization of Ni$^+$ to Ni$^{+2}$ (IP $=$ 18.2\,eV) requires photons  below the Lyman limit. \cmfgen\ models suggest that Ni$^{+2}$ predominates within the inner wind of \ec-A (r $<$ 15 au), but Ni$^{+2}$ absorption  is not easily seen against the continuum of \ec-A. Due to atomic structure of \ion{Ni}{3}, no strong resonant  lines are found above 910\AA.

As the EUV from \ec-B drops near periastron passage, Ni$^{+2}$ in the Little Homunculus and Homunculus recombines, leading to nearly saturated Ni$^{+1}$ absorption from $-$140 to $-$500 \kms.  Near $\phi=11.01$ and $\phi = 14.01$  the absorption profile is saturated between $-400$ and $-500$~\kms\ with a high-velocity blue wing extending to $-$900 \kms\ (Figure~\ref{fig:NiIItrans}). This high velocity absorption wing occurs when the accelerated wind of the primary crosses our LOS as \ec-B moves around \ec-A (Figure \ref{fig:concept}-C). By $\phi = $ 11.119 and $\phi = $ 14.170, the transient absorptions disappear once the EUV radiation re-ionizes the foreground shells and the previously formed cavity. The newly formed cavity  is bounded by a  new downstream bowshock composed of accelerated primary wind that flowed across periastron passage.

\section{Epoch-dependent variations in profiles at similar phase}\label{sec:samephase}

Line profiles near the same phase in different orbital cycles provide key diagnostics that helps understand the
long-term changes in the foreground structures and constrain intrinsic changes to the stars themselves. In this section we constrain non-orbit-related changes in the line profiles by comparing line profiles at similar orbital phases in different orbital cycles.  Because of the complex absorption spectrum arising from intrinsic wind and foreground structures, we can examine the behavior of doublet lines to constrain effects of blending which can effect the shapes of the absorption lines.  We also compare the observed spectra to model spectra to further disentangle the observed profiles.


\subsection{A comparison of three epochs in the early high-ionization state}\label{sec:early}
\begin{figure*}
\includegraphics[width=16.5cm]{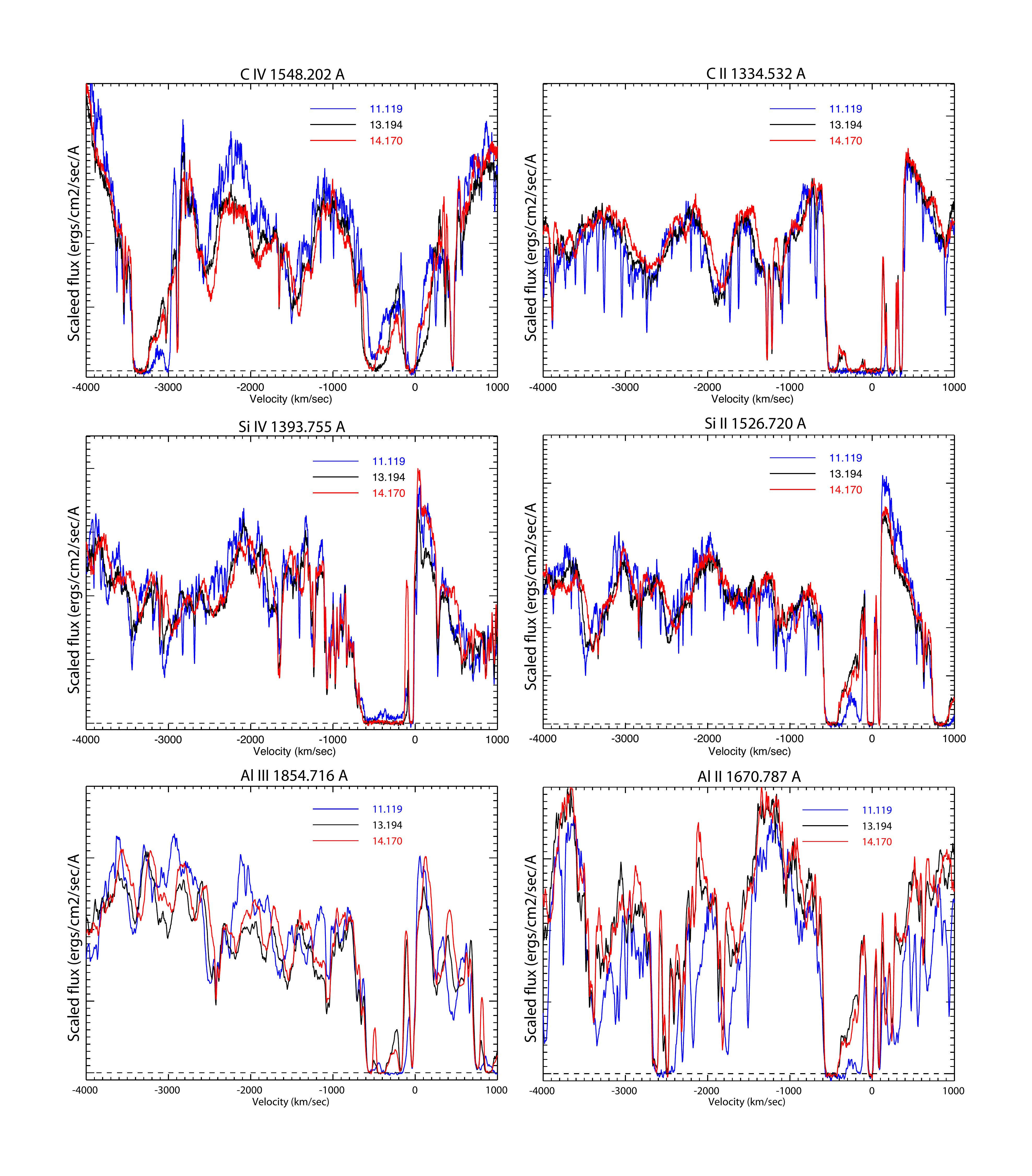}
\caption{Normalized velocity plots of six strong resonant lines recorded at similar orbital phases in the early recovery to  the high-ionization state.  The \ion{C}{4} and \ion{C}{2} doublets are separated by only 498 and 284 \kms, resp., in velocity space, which leads to significant overlap in the 0 to $-$500 \kms\ velocity range of interest. \ion{C}{2}, \ion{Si}{2} and \ion{Al}{2}  reveal decreased absorption after periastron 13 due to the increased FUV radiation leading to multiply-ionized shells within the Homunculus and the Little Homunculus. Lesser changes in absorption are apparent in \ion{Al}{3}, but a  long-term drop in absorption appears to occur at about $-$500 \kms. The \ion{C}{4} $\lambda\lambda$ 1548,1551  absorptions appear to  broaden by $\phi =$ 13.194 then narrow at $\phi =$ 14.170, centered on  $-$500 \kms.  This  suggests that the colliding wind structure in the LOS changes with time in an unpredictable manner. The spectra were normalized in flux relative to the continuum in the 1475 to 1490 \AA\ spectral interval, where relatively few absorption features are apparent. Normalization factors were {\color{blue}  8.25$\times$ ($\phi=11.119$}, 1.54$\times$ ($\phi=13.194$ and {\color{red}  1.00$\times$ ($\phi=14.170$)}. 
} 
\label{fig:Long}
\end{figure*}

A fundamental question is whether the binary winds have changed substantially across the interval sampled by
\hst/STIS observations. Any such changes would indicate intrinsic variations in \ec-A and/or \ec-B.  To address this question,  we compare wind profiles at three similar orbital phases in three different orbital cycles; $\phi =$\ 11.119, 13.194 and 14.170. At these phases the ionization  of the wind structures has almost fully recovered to the high-ionization state near apastron, and the influence of \ec-B 
on the inner primary wind is minimal \citep{Hillier01,Hillier06}. Figure \ref{fig:Long} compares the profiles of six resonance lines which have been normalized  relative to the ``continuum'' at $\phi=14.170$ in the 1475 to 1490\AA\ region, a wavelength region which has few absorption features.  These normalized profiles are astonishingly similar, despite the very different flux scaling factors: {\color{blue} 8.25$\times$ ($\phi =$ 11.119)},  1.54$\times$ ($\phi = 13.194$) and {\color{red}  1.00$\times$ ($\phi=14.170$)}. 
Closer examination of the FUV spectra shows that the narrow line-absorptions of H$_2$  ($-$513 \kms) greatly weakened between 11.119 and 13.194, and are nearly absent by 14.170 (Section \ref{sec:windlines}). Most narrow, low-ionization absorption lines over the velocity interval from $-$120 to $-$180 \kms\ have also weakened or disappeared. These changes were caused by the nearly tenfold increase in FUV flux on the foreground absorbers. 
Fewer changes in the singly-ionized metal absorptions occurred blueward of $-$400 \kms\ but H$_2$ at $-$513 \kms\ was destroyed. By $\phi = 14.170$, EUV radiation further ionized the Little Homunculus ($-146$ \kms), but has not yet strongly affected the Homunculus shell ($-$513 \kms), and the increase in radiation longward of 912\AA\ caused the  photo-destruction of  H$_2$ in our LOS in the most recent spectra. 

The bulk of  changes in broad absorption occurred in the $-$120 to $-$400 \kms\ velocity range.  Absorption in \ion{Al}{2}, \ion{Al}{3}, \ion{Si}{2}  decreased with time most strongly between $\phi =$ 11.119 and 13.194 across that velocity interval. These changes are due to enhanced ionizing flux which indicates that the affected structures reside within or beyond the occulter relative to the position of \ec\ in our LOS.  Some decrease in absorption is present in the \ion{C}{2} profile. However, the \ion{C}{2} doublet overlaps by 1.17\AA\ ($V=284$~\kms), which complicates the tracking of detailed changes in absorption by the individual \ion{C}{2} lines.  
 The blue component of the \ion{Si}{4} doublet ($\lambda$ 1394) shows a P~Cygni profile which is similar at all three phases. However at $\phi=11.119$, the strong absorption  in the $-120 <V< 600$~\kms\ range is not saturated, but is saturated at the two following phases.

The  overlap of the \ion{C}{4}\ doublet (1.59\AA\ or 498~\kms) is comparable to the velocity range of interest and
 possibly contaminated by other species (see Section \ref{sec:wind}), so the profile is challenging to interpret.  At phase 11.119, for example, the the blue component is unsaturated while the red component, which has the lower $gf$ value, appears to be  saturated. However, with knowledge of the intervening absorbing structures, contributions can be parsed out of these overlapping velocity intervals. At $\phi =$ 11.119, the absorptions of $\lambda\lambda$1548,1550  at $-$500 \kms\ are strong, moderately wide, but not saturated. At $\phi =$13.194, the $\lambda$1548 absorption broadened relative to the $\lambda$1550 absorption absorption and relative to the $\lambda$1548 absorption at $\phi =$ 11.119. By $\phi =$ 14.170, both $\lambda\lambda$1548, 1550 profiles  again narrowed and had similar red shoulders.
 
\subsection{Comparing doublet  profiles}\label{sec:doublet}

\begin{figure*}
\includegraphics[width=16.8 cm]{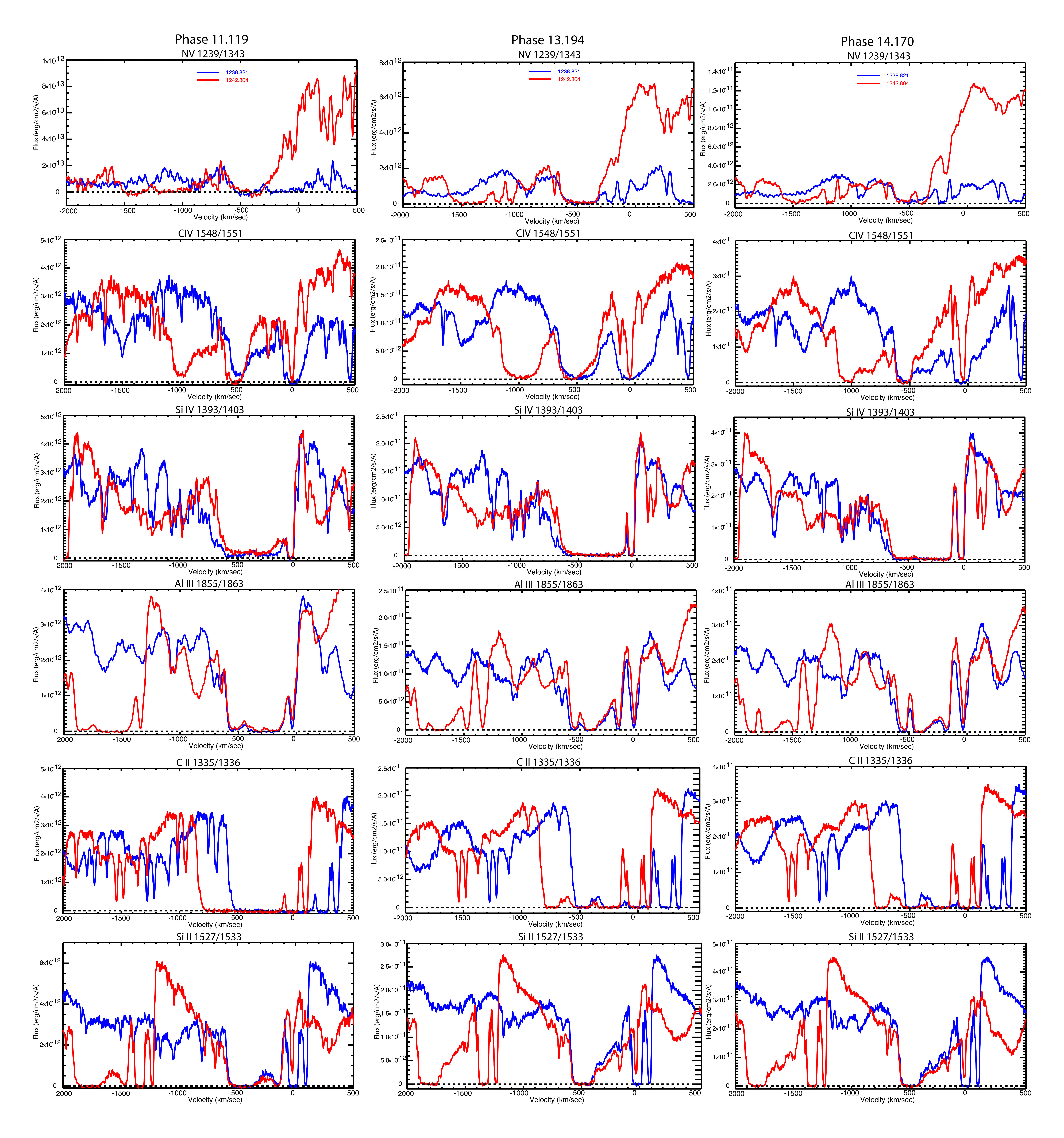}
\caption{Plots of six doublet profiles recorded in three separate cycles near $\phi =$ 0.15 when the winds and multiple Homunculus/Little Homunculus shells are in early recovery to the high-ionization state. The {\color{blue} bluer member of the
doublet is shown in blue} and the {\color{red} longer wavelength member is shown in red}. For \ion{Si}{2} and \ion{Al}{3} the absorption between $-$100 and $-$600~\kms\ is very similar in both members of the doublet, and indicates identical maximum absorption velocities of $\sim$600\,\kms. The \ion{Si}{4} profiles, while similar, indicate somewhat higher velocities. A peculiar notch
at $\sim-$700\,\kms\ is present in the \ion{Si}{4} profiles at all three phases in both doublet members, so must be real. As the longer  doublet member absorption ({\red red}) is consistently weaker than the shorter doublet member ({\color {blue} blue}), the gas causing this absorption is not optically thick. The \ion{C}{4}\ and \ion{N}{5}\ differ from their low ionization counterparts, but the doublet components show similar (although not identical) absorption structure. A simple comparison of the \ion{C}{2} profiles cannot be made because of the small separation ($-280\,$\kms) between the two strongest members. The $+$87\,\kms\ absorption as seen in the shorter wavelength line for \ion{Si}{2}\ and \ion{C}{2}\ is from an extensive IS cloud well in front of \ec\ and, as documented by \cite{Walborn02}, found to be present in the spectra of multiple stars within the Carina region.}
\label{fig:Doublet}
\end{figure*}

By comparing doublet line-profiles  we can disentangle absorption contributions from other lines to obtain  information about absorption optical depths and  covering factors. In the optically thin case the absorption depths of the doublet members  scale as the $gf$-value (Table \ref{tab:lines}), but at large optical depths both lines become saturated. If two lines with different $gf$-values show identical non-zero absorption profiles, the covering factor is likely less than unity (assuming a negligible line source function). 

Figure \ref{fig:Doublet} displays profiles of six strong doublets for the same three spectra recorded in early recovery to the high ionization state  discussed in Section \ref{sec:early}. As noted by \cite{Hillier01, Hillier06}, the influence of \ec-B on the atmosphere and inner wind of \ec-A  is minimal (from our perspective) in the early recovery phase. 
In the three orbital phases shown in Figure \ref{fig:Doublet}, the doublet profiles from 0 to $-$600~\kms\ are nearly identical for \ion{Si}{2} and  \ion{Al}{3}. However, there are important differences for the two species. The  \ion{Al}{3} profiles show an absorption reversal near $-$500~\kms\ at $\phi =$\,13.194, and especially at $\phi =$\,14.170. This reversal is larger for the red component, and is  consistent with the red component having the smaller  $gf$-value. Since a ``normal" wind would give rise to a continuous P\,Cygni absorption profile, the higher velocity \ion{Al}{3} does not originate in the unperturbed wind of \ec-A. A separate absorber gives rise to the  \ion{Al}{3} centered at $-$550 \kms. By contrast, the \ion{Si}{2}\ profiles are nearly black from $-$400 to $-$600 km/s especially at the two later orbits. The profile from $-400<V<-100$~\kms\ is similar in the two profiles, and is broadly consistent with that expected for resonance scattering in a stellar wind plus shell absorptions at $\approx-$150\,\kms\ (analogous  to the \ion{Ni}{2} $\lambda$1455 velocity profile seen in Figure \ref{fig:NiII10to13}).
By contrast, the profiles of the \ion{C}{2} doublet are very different between $-$100 and $-$400 \kms.  The ({\red red}) $\lambda$1336 absorption is noticeably weaker than the ({\color {blue} blue}) $\lambda$1335 absorption, again consistent with the $gf$-values.

The \ion{Si}{4} doublet absorption profile at $\phi =$ 11.119 is nearly saturated and the red component of the doublet  shows a larger residual flux from $-100<V<-600$~\kms.  This is probably due to the extended wind of \ec-A, which dominates the total flux at $\phi=$ 11.119, compared to later times when the occulter dissipated and continuum from the core of \ec-A  dominates.  At $\phi =$ 13.194 and 14.170, both profiles are saturated. This saturation is especially remarkable, since it means that there is highly ionized gas that covers a broad velocity range along our line of sight, and which covers the entire continuum-emitting region of the binary system. The distribution of gas which produces the broad \ion{Si}{4} P-Cygni absorption must be asymmetric, since if it were symmetric, absorption from a normal, unperturbed stellar wind, the P~Cygni profile would more closely resemble the \cmfgen\ model profile of \ion{Si}{2}\ $\lambda1533$ (Figure \ref{fig:CMFGEN}).

The blue-edge velocities indicate the maximum speed of
the gas along our line of sight. The blue-edge profiles of \ion{Si}{2} and for \ion{Al}{3} are very similar, saturated  to $V\approx-$550\,\kms. The profile is black only to a velocity of $-$520\,\kms, but the \ion{C}{2} profile is complicated by the extreme velocity overlap of the doublet. The \ion{Si}{4} blue edge on the blue component of the doublet has a steeper slope than the red component, and might be attributable to optical depth effects (the blue component has a $gf$ value that is a factor of 2 larger) plus incomplete coverage of the source.

The \ion{C}{4} doublet velocity-profiles show considerable differences in the three observations. At $\phi =$ 11.119 both lines show deep absorption extending from  about $-$500 to $-$600 \kms. However the redder component of the doublet shows evidence for extra absorption by another species, since it is deeper, and extends to slightly higher velocities than the 1548 component, despite its $gf$-value being  a factor of 2 lower than the blue component. Comparisons with the \cmfgen\ model in Section\,\ref{sec:CMFGENmod} show that  iron absorption in the spectrum of \ec-A contaminates the red doublet component. At $\phi =$ 13.194, both \ion{C}{4} lines show strong, broadened absorption extending from $-$300 to $-$650 \kms, much broader than that seen at phases 11.119 and 14.170. At $\phi=14.170$, the lines again show deep absorption extending from  about $-$500 to $-$600~\kms, but the edge velocities are identical. Both components show absorption between $-$100 and $-$500~\kms\ that is broadly consistent with the $gf$-values, although confusion caused by line overlap could play a role. The  broad component of $\lambda$1550 obscures the $-$40 \kms\ component of $\lambda$1548. But the $\lambda$1550 $-$40 \kms\  narrow component is present, confirming strong, well-defined absorption at that velocity, which is characteristic of the Weigelt clumps B, C and D \citep{Gull16}. 

We confirm the presence of the \ion{N}{5} doublet from the  broad, nearly saturated absorption extending from $-$300 to $-$600 \kms\ in both doublet components (velocity separation of 915 \kms) in all three observations. However, the profile is noticeably broader at $\phi =$ 11.119 than at 13.194 and 14.170. As noted by \cite{Gull21a}, the wavelength-dependent absorption steepens below 1300\AA, so the \ion{N}{5} profiles were normalized at 1250\AA\ for comparison. 

\subsection{Comparing observed and model profiles}\label{sec:wind}

We now compare early recovery profiles just after periastron passage to model profiles  of \ec-A  from \cmfgen\ (Section \ref{sec:CMFGENmod}) to estimate the contribution of the primary wind to the observed profiles, and to identify possible contributions from \ec-B, the colliding winds and the discrete foreground shells within the Homunculus. 

\begin{figure*}
\centering
\includegraphics[width=18cm]{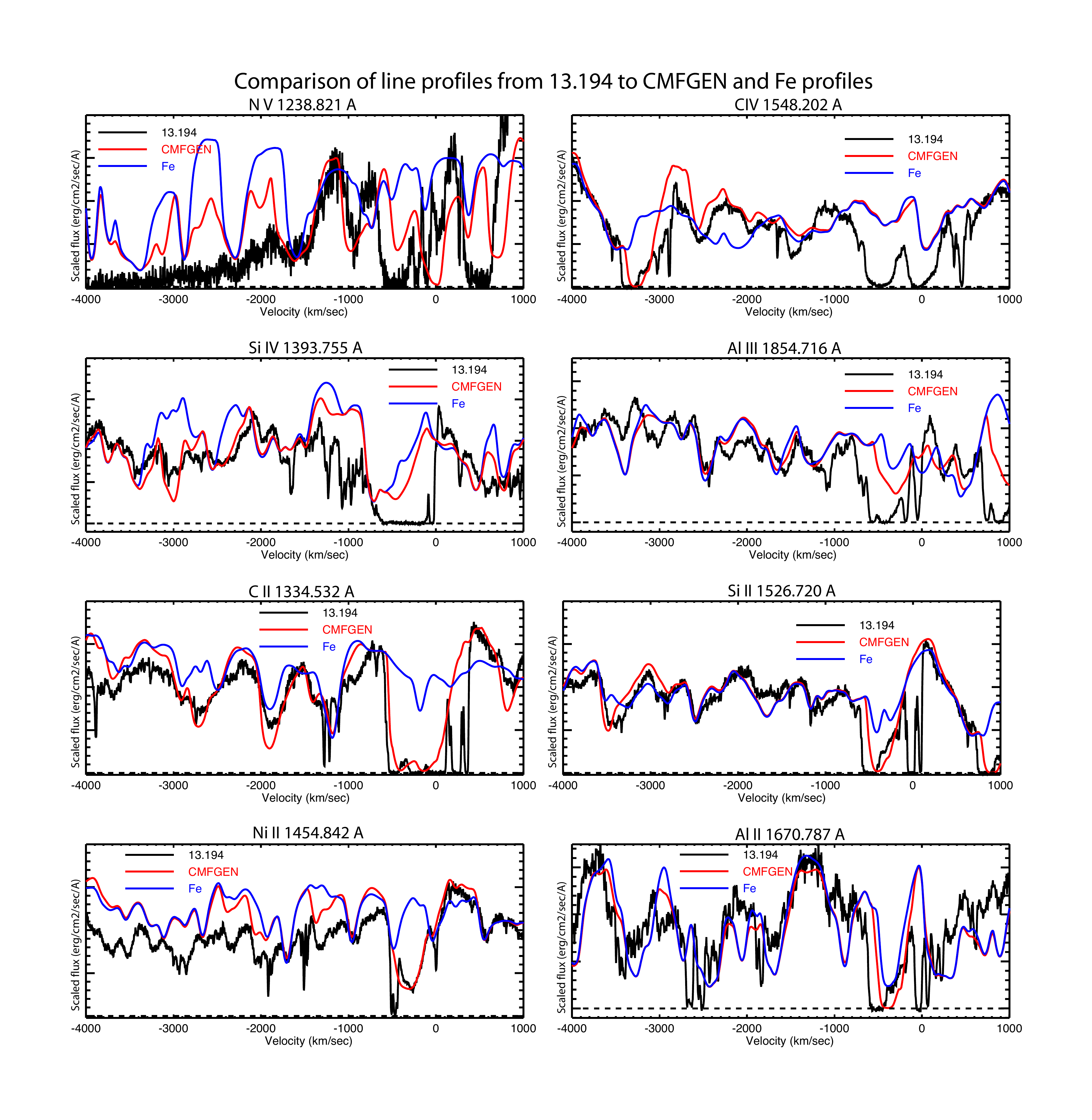}
\caption{ Comparison of \cmfgen\ model profiles with observed profiles recorded at $\phi =$ 13.194. The \cmfgen\ model of \ec-A, which used \Vinf$_{,A}$= 420 \kms, accounts for the bulk of the 0 to $-$500 \kms\ absorption for the low-ionization species up to \ion{C}{2}, less so for \ion{Al}{3} and \ion{Si}{4} and little or no  absorption  by \ion{C}{4} and \ion{N}{5}.
} 
\label{fig:CMFGEN}
\end{figure*}
\begin{figure*}
\centering
\includegraphics[width=12.5cm]{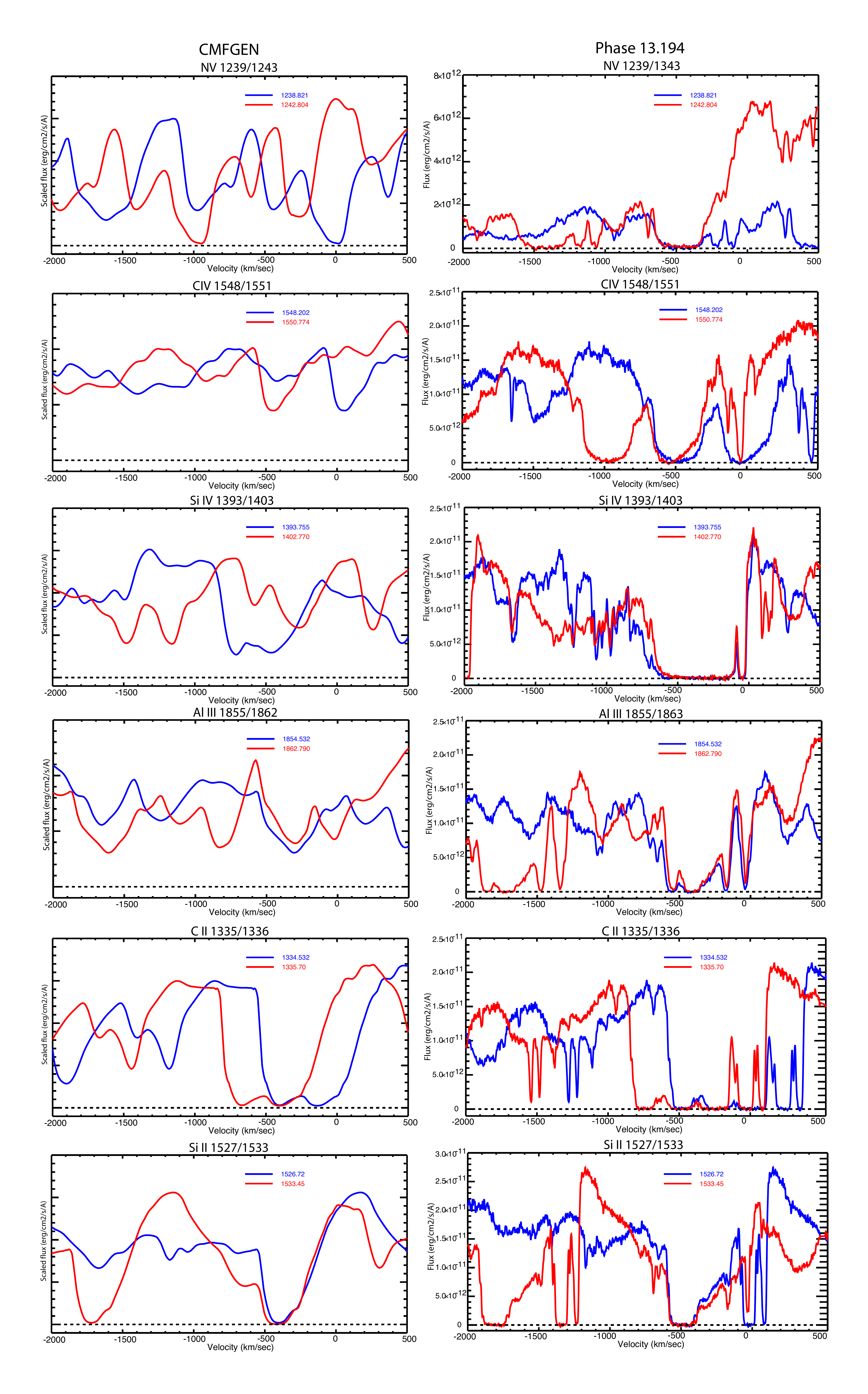}
\caption{ Comparison in velocity space of six doublet pairs from the CMFGEN mode (with \Vinf$_{,A} = 420$ \kms, left column) and  observed doublet pairs from the $\phi =$ 13.194 spectrum (right column).
In the model, \ec-A contributes a significant wind absorption  between 0 to $-$500 \kms\ for low ionization lines from \ion{Si}{2} through \ion{Al}{3}, little absorption for the \ion{Si}{4} and negligible absorption for \ion{C}{4} or \ion{N}{5}. {\color{blue} The shorter wavelength line of the doublet is plotted in blue} and {\color{red} longer wavelength line is plotted in red.}} 
\label{fig:CMFGENDoublet}
\end{figure*}
Figure~\ref{fig:CMFGEN} compares profiles at $\phi =$ 13.194 with  \cmfgen\ model profiles (shown in {\red red}), along with the expected contribution of iron to the spectrum (in {\color{blue} blue}). Isolating the iron spectrum helps identify most of the contaminating primary wind absorptions near the resonant lines of interest. As shown in Section \ref{sec:NiII}, the \cmfgen\ model spectrum produces a velocity profile for \ion{Ni}{2} that is very similar to the observed velocity profile (Figure \ref{fig:CMFGEN} lower left), and helps identify the $+$4 and $-$500 \kms\ absorptions as arising from  an interstellar cloud and the Homunculus, respectively (see Figure \ref{fig:NiII10to13}).
In general, the models show a significant primary wind contribution
to low-ionization lines (\ion{C}{2}, \ion{Al}{2}, \ion{Si}{2}), but little or no contribution to the high-ionization lines (\ion{Si}{4}, \ion{C}{4}, \ion{N}{5}). For the low-ionization species, the \cmfgen\ profiles extend to about $\sim -$420 \kms\ as expected for the primary wind; however
the observed profiles are saturated out to $V\sim -550$\,\kms.

Figure \ref{fig:CMFGENDoublet} compares the doublet profiles from the \cmfgen\ model  with the observed profiles. The observed doublet profiles are quite similar  from \ion{N}{5} (IP $=$ 77.47 eV) through \ion{Si}{2} (IP $=$ 8.15 eV), which reinforces the idea that the absorptions originate from the specific resonant lines.  However, the doublet profiles from the \cmfgen\ model are similar only for the low-ionization lines (\ion{Si}{2}, \ion{C}{2}). The \cmfgen\ model also shows weak \ion{Al}{3} P~Cygni absorption. The \cmfgen\ profile of the  \ion{Al}{3}\  doublet may be contaminated by  iron absorption which could be due to the influence of \ec-B on the  outer wind of the \ec-A.

\subsection {Long-term changes} \label{sec:longterm}

To constrain long-term (non-orbit-related) changes in the system, we compare  spectra at nearly the same phase  in the velocity range of $-$4000 to $+$1000 \kms\ for select lines (Figures  \ref{fig:C4C2} through  \ref{fig:Al3Al2}).  Following  \cite{Groh10} , for our comparison, we normalized the spectra to  the continuum near 1487\AA\ ($\lambda\lambda$1475 to 1490). However, we used the continuum near 1800\AA\ ($\lambda\lambda$1775 to 1805) when scaling the aluminum doublet as the observations were done with the E230M echelle.

\subsubsection{Comparison of \texorpdfstring{\ion{C}{4}}{CIV} and \texorpdfstring{\ion{C}{2}}{CII} resonant line profiles}{\label{sec:C4C2}}

\begin{figure*}
\includegraphics[width=18.cm]{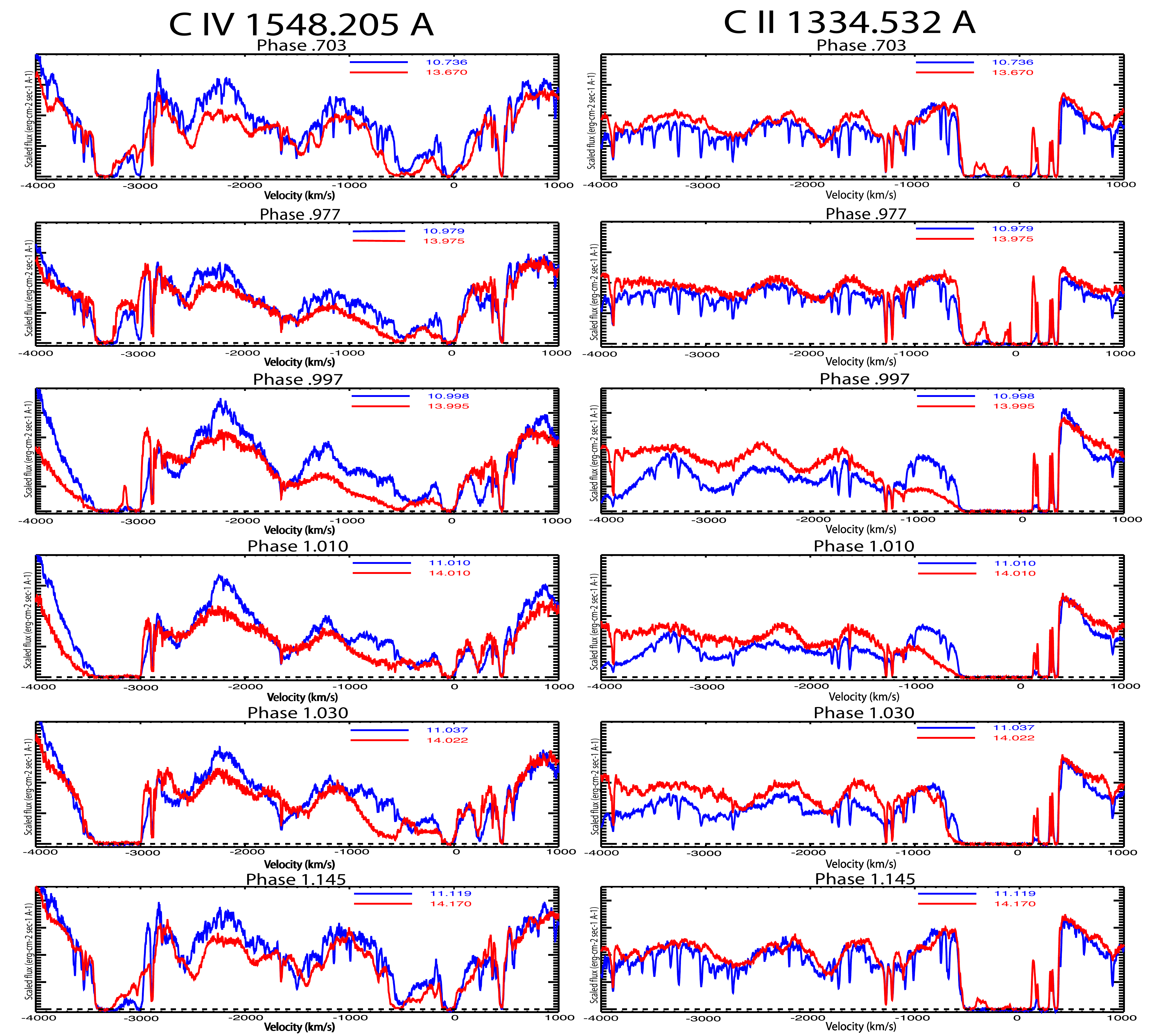}
\caption{ Comparison of \ion{C}{4} (IP $=$ 47.9/64.5 eV)  and \ion{C}{2} (IP $=$ 11.3/24.4 eV) profiles at six similar phases across cycles 10/11 ({\color{blue} periastron passage 11}) and 13/14 ({\color{red} periastron passage 14}). {\bf Left Column:}  The difference in \ion{C}{4} $\lambda$1548 absorption profiles indicates increased absorption from 0 to $-$2600 \kms\ by cycle 13 compared to Cycle 10. {\bf Right column:}  \ion{C}{2} $\lambda$1335 absorption profiles  reveal a ramp-like absorption extending from $-$100 to $-$600 \kms\ across periastron passage 14 that was not present across periastron passage 11. Flux scaling was based on continuum levels from 1463 to 1488\AA.}
\label{fig:C4C2}
\end{figure*}

Figure \ref{fig:C4C2} compares \ion{C}{4} $\lambda$1548 (IP $=$ 64.5/392 eV) and \ion{C}{2} $\lambda$1335 (IP $=$ 11.3/24.4 eV) at nearly the same orbital phases from cycles 10/11 and 13/14. The normalized \ion{C}{2} profiles leading up to periastron passages 11 and 14 ($\phi =$ .703 to .977) are surprisingly similar throughout the  entire velocity range $-600<V<-4000$~\kms,and in the early recovery to high-ionization state after periastron passages ($\phi =$1.145). Near  the periastron passage in orbital cycle 14 ($\phi = 13.995$ \& $14.010$), the \ion{C}{2} profile shows a blue absorption  wing extending from $-1000<V< -600$~\kms\  that was not present during periastron passage 11.  This wing largely disappeared by $\phi = 14.022$  and fully disappeared by $\phi = 14.170$, when once again there was good agreement between the normalized line profiles from cycle 14 and cycle 11.

Because the velocity spacing between \ion{C}{2} $\lambda$1335 and $\lambda$1336 is only 284~\kms, the spectral component are blended which complicates the comparison of  \ion{C}{2} with \ion{C}{4} $\lambda\lambda$1548,1555. Small ``continuum" bumps visible between 0 and $-$600 \kms\ during the high-ionization state disappear near periastron passages in both cycles. This change is likely due to a change in ionization -- the carbon in some shells is multiply-ionized during the high ionization state but returns to a less-ionized state near periastron passage. 

The blue absorption wing that appeared in \ion{C}{2} near periastron passage 14 is very similar to the wing seen in three \ion{Si}{2} doublet profiles shown in \cite{Gull21a}. They suggested that the repetition of this blue absorption wing in  \ion{Si}{2} is likely caused by dense clumps of material in the the trailing arm of the bow shock which move into our LOS during periastron passage.  \cite{Pittard07} modeled interactions of clumps with wind-wind shocks and determined that the clumps are rapidly destroyed. The transient blue-absorption wing we see for about 70 days around periastron passage  may be an example of clump destruction in the bow shock on this timescale.

The changes seen in the \ion{C}{4} profile are similar to those seen in the \ion{C}{2} line.  The \ion{C}{4} line also  shows a blue  absorption wing near periastron in cycle 14 at the same phases where  the blue wing is seen in the \ion{C}{2} line. The \ion{C}{4} blue wing in cycle 14 extends  from velocities between $-100<V<-1400$\,\kms, and probably to $-$2600~\kms\ (which is near the expected terminal velocity of the wind of \ec-B). \cite{Groh10}, using the same \hst/STIS spectra recorded before  periastron passage 11, identified a transient absorption extending to $-$1200 \kms, and also showed transient high-velocity absorption in \ion{He}{1} $\lambda$10830 extending to $-$1900 \kms. The \ion{C}{4}\ absorption extending to $-$2600 \kms\ suggests that  our LOS extends deeper into the wind-wind structure due to the disappearing occulter.  This is in line with the greatly increased FUV flux and the apparent decrease of the H$\alpha$\ equivalent width over the past few decades \citep{Mehner12,Mehner15, Damineli21}. However, the extended blue absorption wing seen in  \ion{C}{4} may be blended with other variable absorptions near periastron passage, especially at the most blueward velocities. In contrast to \ion{C}{2}, high velocity absorption is also present in \ion{C}{4}  from $-$100 to$-$2600 \kms\ at $\phi =$ 0.703 in cycle 14. The absorption appears to be in two velocity intervals, from $-$100 to $-$1500  \kms\ and then between $-$2000 to $-$2600 \kms.  This suggests that both the \ion{C}{4} $\lambda$1548 and $\lambda$1551 lines are absorbing in the $-$2000 to $-$2600 \kms\ velocity range in addition to the  $-$500 to $-$600 \kms\ interval. 

\subsubsection{Comparison of \texorpdfstring{\ion{Si}{4}}{SiIV} and \texorpdfstring{\ion{Si}{2}}{SiII}  Doublet Absorptions} \label{sec:S4S2}

\begin{figure*}
\includegraphics[width=18.cm]{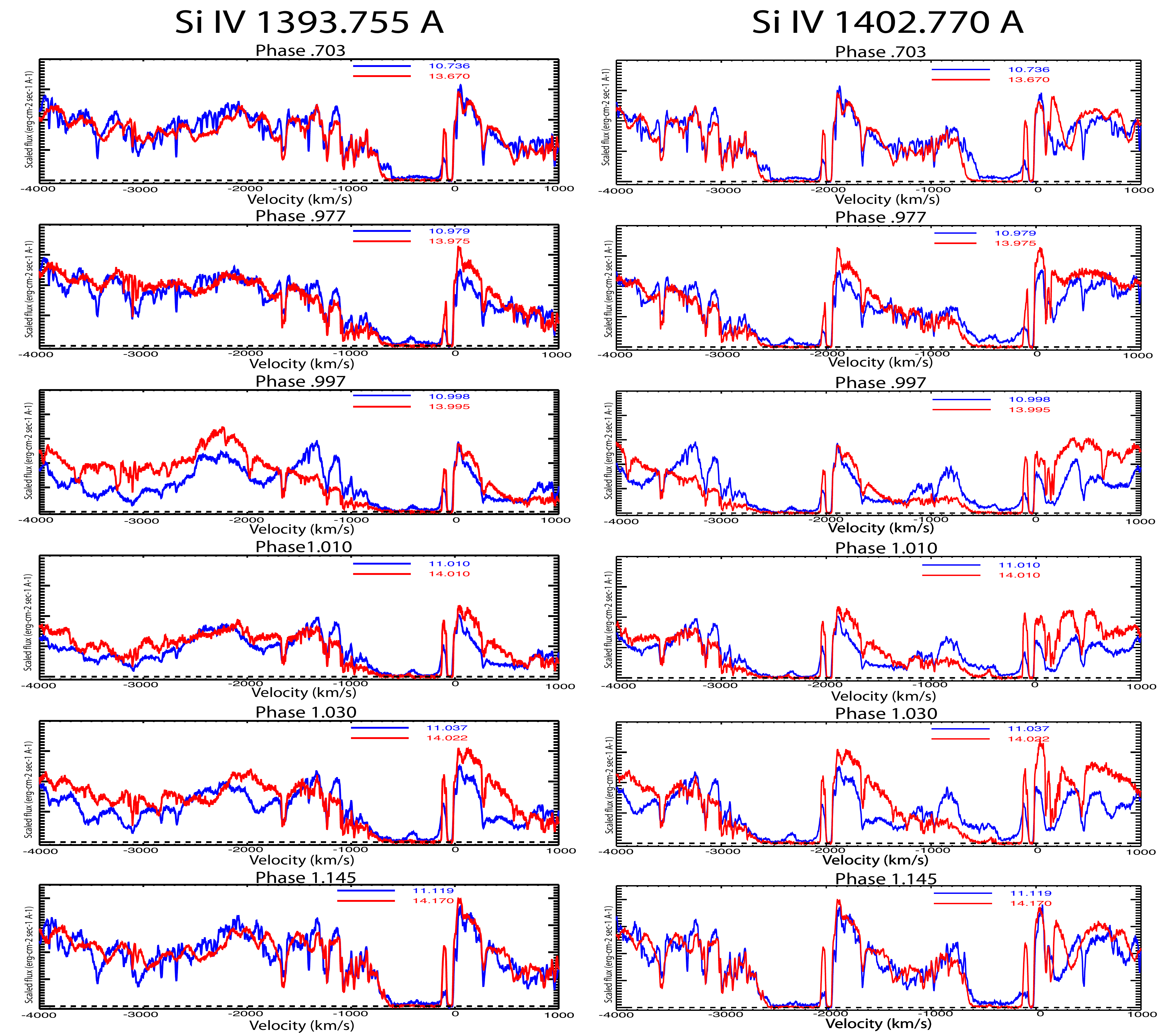}
\caption{Comparison of \ion{Si}{4} (IP $=$ 45.1 eV)  at six similar phases across cycles 10/11 ({\color {blue} periastron passage 11}) and 13/14 ({\color{red}peristron passage 14}). Although the \ion{Si}{4} lines show similar absorption profiles leading up to periastron passage, the absorption at $\phi=13.6$ extends further to the blue by about 100\,\kms\ -- the difference is real since it is seen in both components. After periastron passage the profiles are again very similar. The change in the ``spike" near 100\,\kms\ is likely due to changes in the circumstellar medium. Further examination shows that \ion{Si}{4} profiles during cycle 10 show a larger residual intensity in the absorption trough than during cycles 13/14. Significant differences between $-$1000 and $-$2500 \kms\ occur as periastron passage  begins ($\phi =$ 0.997) -- the extended absorption is more obvious, is seen in both doublets, and appears to extend to velocities beyond -2000\,\kms\ in the cycle 13/14 data.}
\label{fig:Si4}
\end{figure*}

\begin{figure*}
\includegraphics[width=18.cm]{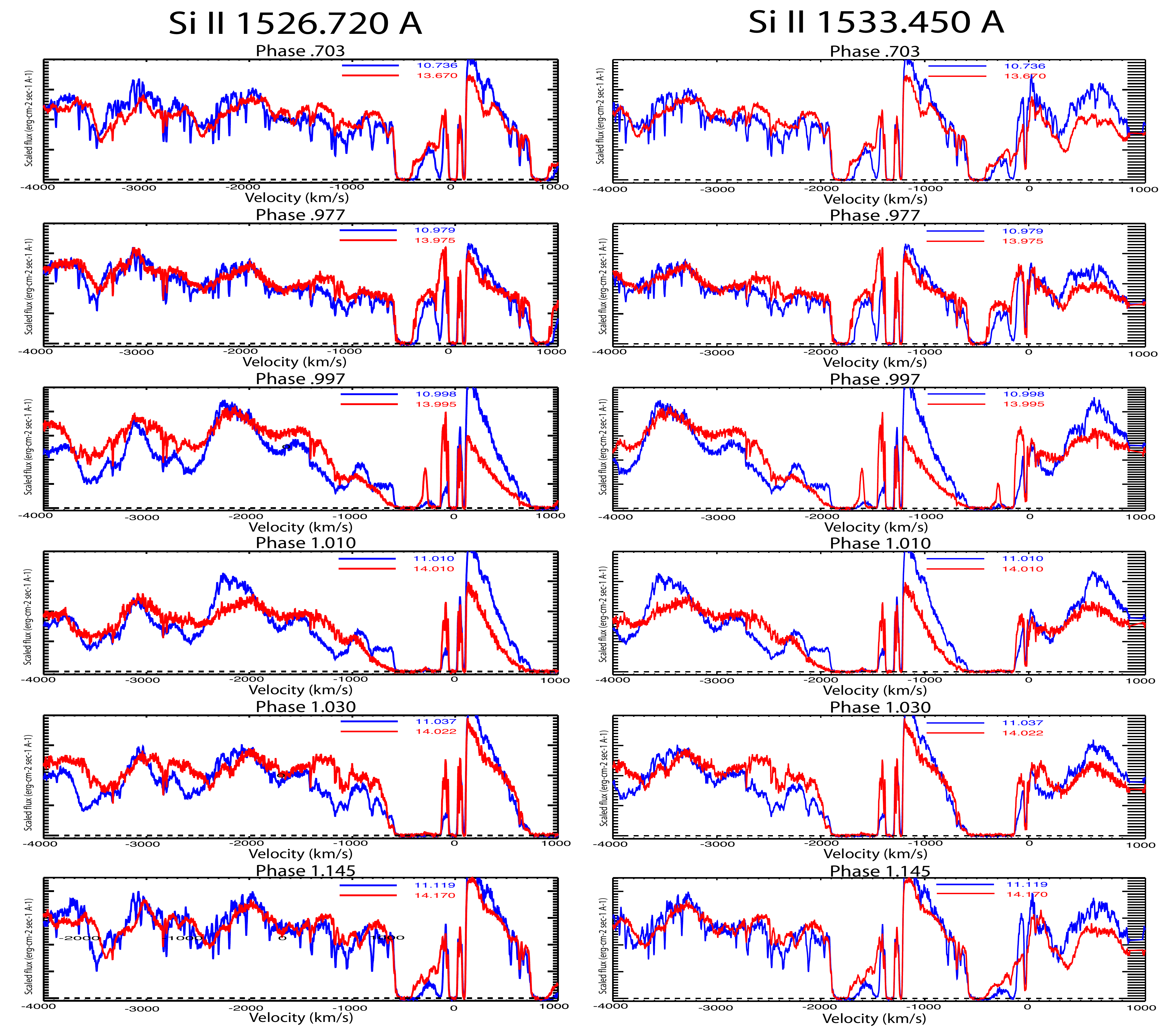}
\caption{ Comparison of  \ion{Si}{2} (IP $=$ 8.2 eV) profiles at six similar phases across cycles 10/11 ( {\color{blue}periastron passage 11}) and 13/14 ({\color{red} peristron passage 14}). As for \ion{Si}{4} the profiles outside of periastron passage
are very similar, with differences most likely a consequence of blending. The biggest difference is the
absence/weakness of the $-146$\,\kms\ component, which arises in the LH, in the 13/14 cycle. During periastron passage absorption has  increased over $-$600 to $-$1200 \kms, with longer wavelength absorption eating into the P~Cygni emission component of the bluer
member of the doublet.}
\label{fig:Si2}
\end{figure*}

The normalized \ion{Si}{4} (IP $=$ 33.5/45.1 eV)  and \ion{Si}{2} (IP $=$ 8.2/16.4 eV)   profiles are displayed in two separate figures as the doublet lines are separated sufficiently to allow independent comparisons  of the lower velocity range (Figures \ref{fig:Si4} and \ref{fig:Si2}). Indeed the  shorter wavelength members of the two doublets have no competing  strong absorptions up to $-$4000 \kms, well beyond the expected terminal velocity of \ec-B \citep{Pittard02}. 

The  normalized \ion{Si}{4}  profiles (Figure \ref{fig:Si4}) show no strong variations in absorption in the high-ionization phase either well before the periastron passage or in the early recovery to the high-ionization state. The absorption is nearly saturated from $-$100 to $-$600 \kms, especially for the $\lambda$1394 line, but continuum is present for the $\lambda$1403 line in the phases plotted in Figures \ref{fig:Si4} and \ref{fig:Si2}. 
Significant changes in the \ion{Si}{4} doublet lines do occur between  the  periastron passages cycles 11 \& 14.  Most notably, at $\phi =$ 0.997 absorption increased in the velocity range $-1000<v< -1600$~\kms\ in cycle 14 compared to cycle 11. Increased absorption in  \ion{Si}{4} $\lambda$1403 continued throughout that velocity range across periastron passage 14. 


The normalized \ion{Si}{2}  profiles (Figure \ref{fig:Si2}) show stronger absorption in the range  $-100< V <-400$~\kms\ for the $\lambda$1533 line compared to the $\lambda$1527 line,  consistent with the $gf$ values. Completely saturated absorption occurs from $-$400 to $-$600~\kms\ for both lines. A noticeable increase in absorption from $-$800 to $-$1400~\kms\ is present in the $\lambda$1527 line (and up to $-$1200 \kms\ in the $\lambda$1533 line)  in the high-ionization state across periastron passage 14 relative to periastron passage 11. Comparison of the $\lambda$1527 absorption profile to the \cmfgen\ model (Figure \ref{fig:CMFGEN})  indicates that the continuum in the high-ionization state approaches the model velocity profile in the $-$100 to $-$400~\kms\  interval.   By cycle 14, the $-$170 \kms\ absorption from the Little Homunculus disappeared across the high-ionization state because the increased FUV flux has caused most metal ions to become doubly-ionized. Across the periastron passages, when the FUV flux of \ec~B is hidden as it passes behind the primary, the metals return to singly-ionized state and absorption becomes  nearly completely saturated from $-$100 to $-$500 \kms\ (Figure \ref{fig:Si2}). However weak continuum bumps are present at $-$100 and $-$300 \kms\ in both cycles at $\phi =$ 0.997. The $\lambda$1533 line is completely absorbed from $-$40 to $-$500 \kms\ at $\phi =$ 1.010 and 1.030, but a spike of continuum survives at $-$100 \kms\ for the $\lambda$1527 line.  This indicates a gap in velocity between the slow moving structure related to the Weigelt clumps \citep{Gull16} and the Little Homunculus.

\subsubsection {Comparing \texorpdfstring{\ion{Al}{3}}{AlIII} and \texorpdfstring{\ion{Al}{2}}{AlII} profiles}
\begin{figure*}
\includegraphics[width=18.cm]{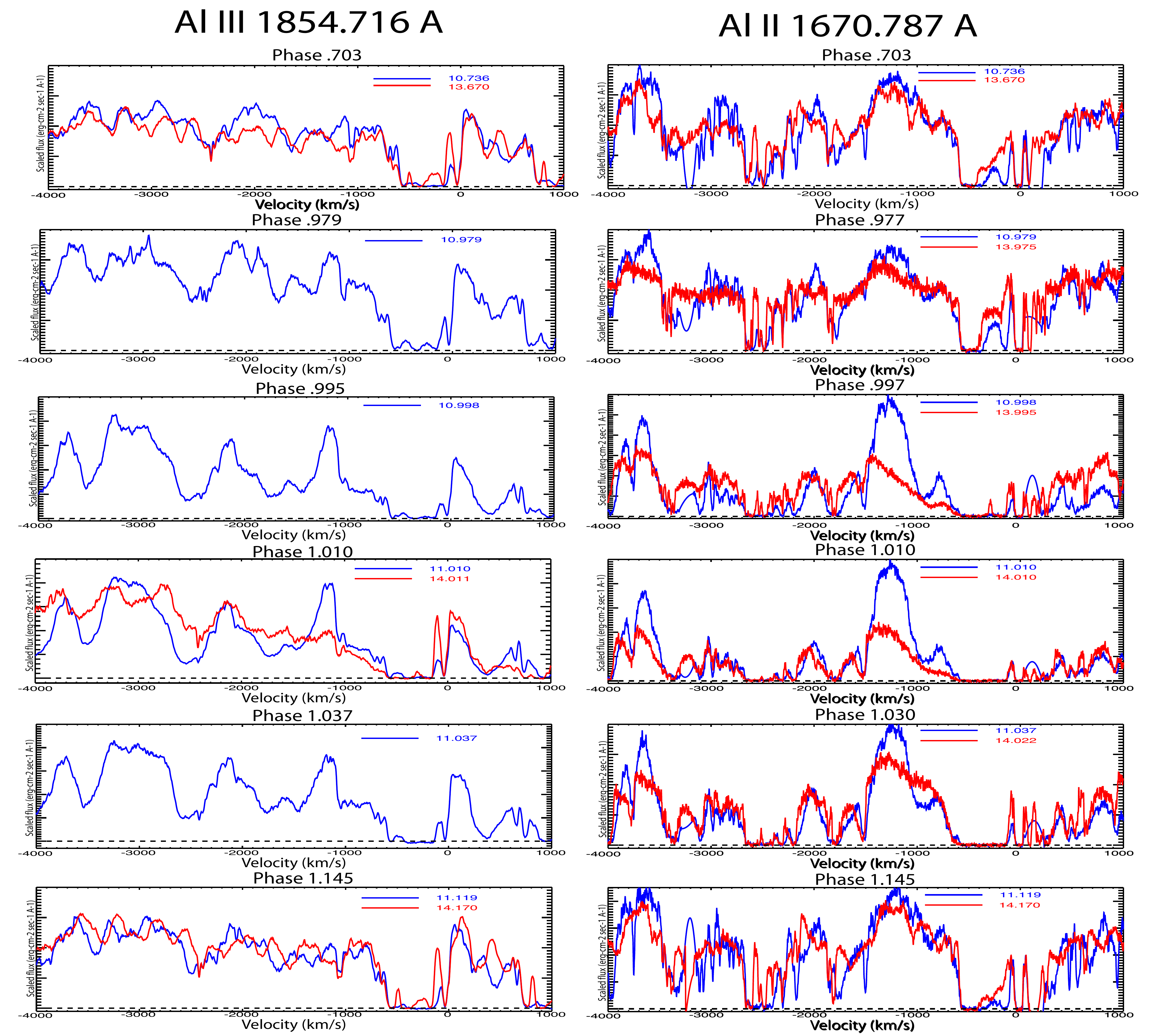}
\caption{ Comparison of \ion{Al}{3} (IP $=$ 18.8/28.5 eV) and \ion{Al}{2} (IP $=$ 6.0/18.8 eV) profiles at six similar phases across cycles 10/11 ({\color {blue}periastron passage 11}) and 13/14 ({\color{red}peristron passage 14}).
{\bf Left column:} Only three comparison spectra are available for \ion{Al}{3} $\lambda$1855 absorption profiles  across cycles  10/11 (periastron passage 11) and 13/14 (periastron passage 14). Notable is the strong absorption in the $-$1000 to $-$1300 \kms interval at $\phi =$ 1.010. {\bf Right column:} \ion{Al}{2} $\lambda$1671 absorption profiles show minimal difference leading up to and after the periastron passages. As with all low ionization lines, a ramp-like absorption appears from $-$600 to $-$!400 \kms across periastron passage 14. Both the \ion{Al}{3} and \ion{Al}{2} profiles were normalized to the continuum near 1800\AA.   Note that the flux is invalid for \ion{Al}{2} from $+$100 to $+$400 \kms\ due to echelle format dropout.} 
\label{fig:Al3Al2}
\end{figure*}

The \ion{Al}{3} (IP $=$ 18.8/28.5 eV) and \ion{Al}{2} (IP $=$ 6.0/18.8 eV) absorptions probe the ionized gas above and below the ionization potential of hydrogen (IP $=$ 13.6 eV), which is  useful for probing the extended primary wind. However, the  \ion{Al}{3} is recorded by the E230M STIS grating which unfortunately was not always used leading up to periastron passage 14. The behavior of the three available profiles of \ion{Al}{3} $\lambda$1855 is similar to that of the \ion{Si}{4} doubles. The profiles in the $\phi =$ 0.703 spectrum and $\phi=1.145$. are quite similar for \ion{Al}{2} and \ion{Al}{3}. At $\phi =$ 1.010 both  lines show high velocity absorption wings extending from $-$1000 to $-$1300 \kms, similar to the high velocity absorption seen in the C and Si profiles.  The profiles are nearly saturated in the $-$100 to $-$600 \kms\ range, but show less saturation in the $-$100 to $-$400 \kms\  range.  The \ion{Al}{2} $\lambda$1671 profiles show little variation in the high-ionization state, $\phi=$ 0.703 and 0.933, but show increased absorption in the  $-$100 to $-$400 \kms\ range from $\phi =$ 0.977 to 1.030.

\subsection{Changes in absorption profiles across the two periastron passages}\label{sec:low}

\begin{figure*}
\includegraphics[width=18.cm]{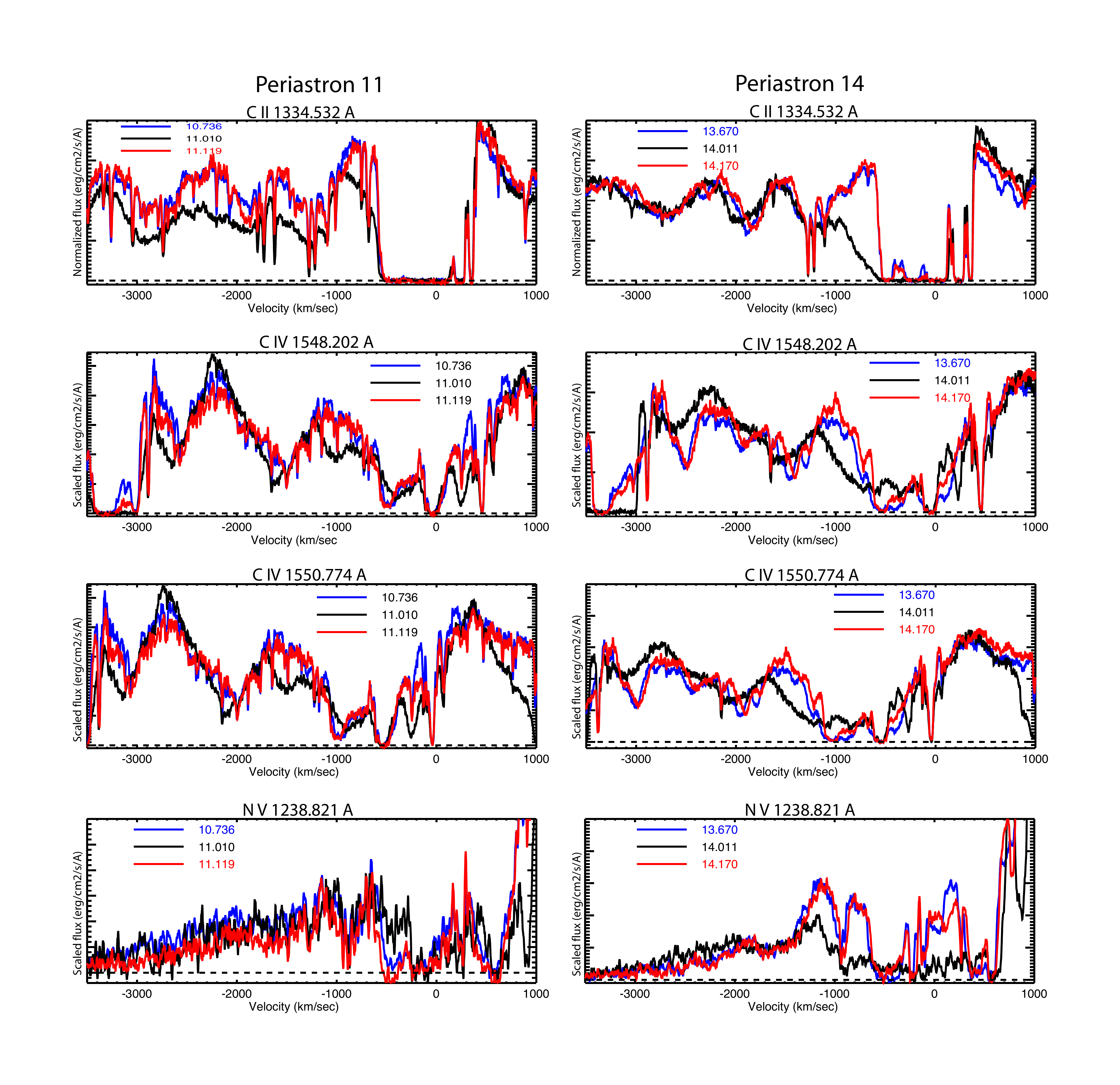}
\caption{ Variations of profiles across periastrons 11 ({\bf Left column}) and 14 ({\bf Right column}).  {\bf Top row:}  \ion{C}{2} $\lambda$1335,  {\bf Middle rows:} \ion{C}{4} $\lambda\lambda$1548, 1551 and  {\bf Bottom row:} \ion{N}{5} $\lambda$1239.  The most noticeable difference is the ramp-like absorption from $-$500 to $-$1200 \kms in the \ion{C}{2} and \ion{C}{4} profiles during  periastron passage 14.  The \ion{C}{2} absorption, which shows islands of lower absorption before and after periastron passage 14, becomes completely saturated during periastron passage 14. While strong variations occur in the \ion{C}{4}$\lambda$1551 velocity profile from $-$100 to $-$3000 \kms, the narrow absorption centered on $-$40 \kms\ does not change. ( \ion{C}{2} (doublet separation equivalent to 284 \kms), \ion{C}{4} (separation 498 \kms) and \ion{N}{5} (915 kms) overlap so closely that complete isolation of doublet contributions is not possible.)}   
\label{fig:C2C4N5P}
\end{figure*}
The onset of periastron led to a plethora of  changes in the STIS FUV spectrum across both periastron passages 11 and 14. From a broadband perspective, the flux declined 50\%, based on the STIS global count rates, then recovered to pre-periastron levels. Most of the flux decline is caused by singly ionized species that become more prevalent along our line of sight due to a drop in ionizing flux near periastron passage when  the shock cone around the companion points away from the foreground gas and the hot companion is buried in the inner wind of \ec-A. Considerable changes occur across both periastron passages. High velocity changes in the resonant line profiles, are stronger near periastron passage 14 compared to periastron passage 11. We do not know whether the observed variations were a continuous trend or part of an abrupt change in the system. However, visible-band ground-based  monitoring observations show a  continuous  flux increase \citep{Damineli21}, which suggests that the changes in the FUV are most likely part of this same continuous change.

\begin{figure*}
\includegraphics[width=18.cm]{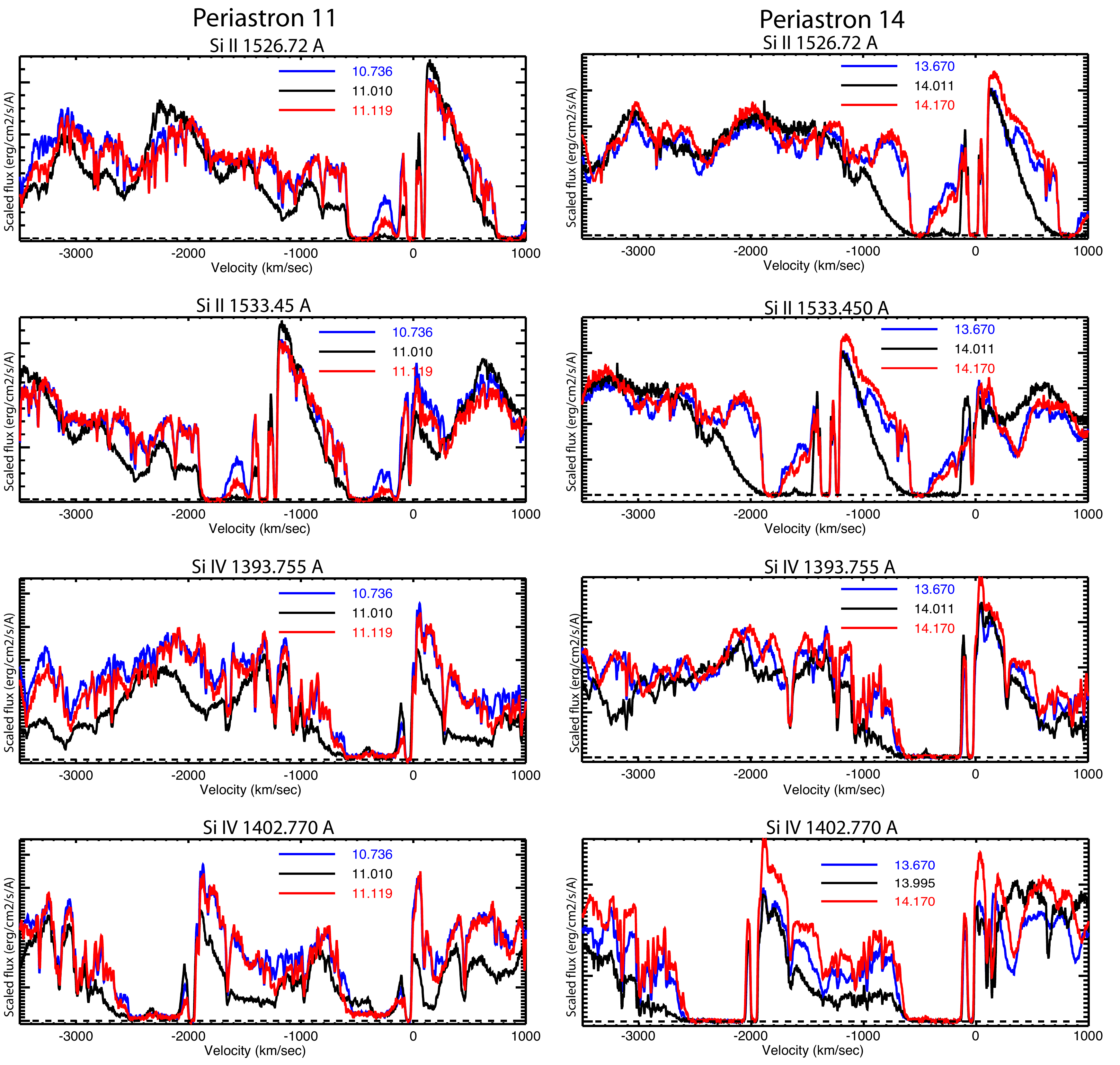}
\caption{ Comparison of   profiles across  periastrons 11 ({\bf Left Column}) and 14 ({\bf Right Column}). {\bf Top two rows:} \ion{S}{2} $\lambda\lambda$1527,1533. The \ion{Si}{2} $\lambda$1527, and possibly the \ion{S}{2}$\lambda$1533, appear to have two components of absorption between $-$500 to $-$700 \kms\ and $-$1000 to $-$1300 \kms across periastron 11 while both \ion{Si}{2} lines show a very smooth absorption ramp  from $-$500 to $-$1000 \kms\ across periastron passage 14.  Two islands of lesser absorption are present before and after periastron passage 11 centered on $-$100 and $-$300 \kms, that disappear across periastron passage 11. The lesser absorption extends from $-$100 to $-$350 \kms\ before and after periastron passage 14, but disappears across the passage. {\bf Bottom two rows:} \ion{Si}{4} $\lambda\lambda$1394,1403. The \ion{S}{4} profiles are affected by other absorption lines. Across periastron passage 11, the $-$100 to $-$600 \kms\ region is not completely saturated, indicating a drop in FUV ionizing flux, but both before and after periastron passage the profiles are quite similar. The \ion{Si}{4}$ \lambda$1403 shows an increased saturation from $-$1300  to $-$1900 \kms. Across periastron passage 14, the $-$100 to $-$600\ \kms\ velocity interval is completely saturated.} 
\label{fig:Si2Si4P}
\end{figure*}

\begin{figure*}
\includegraphics[width=18.cm]{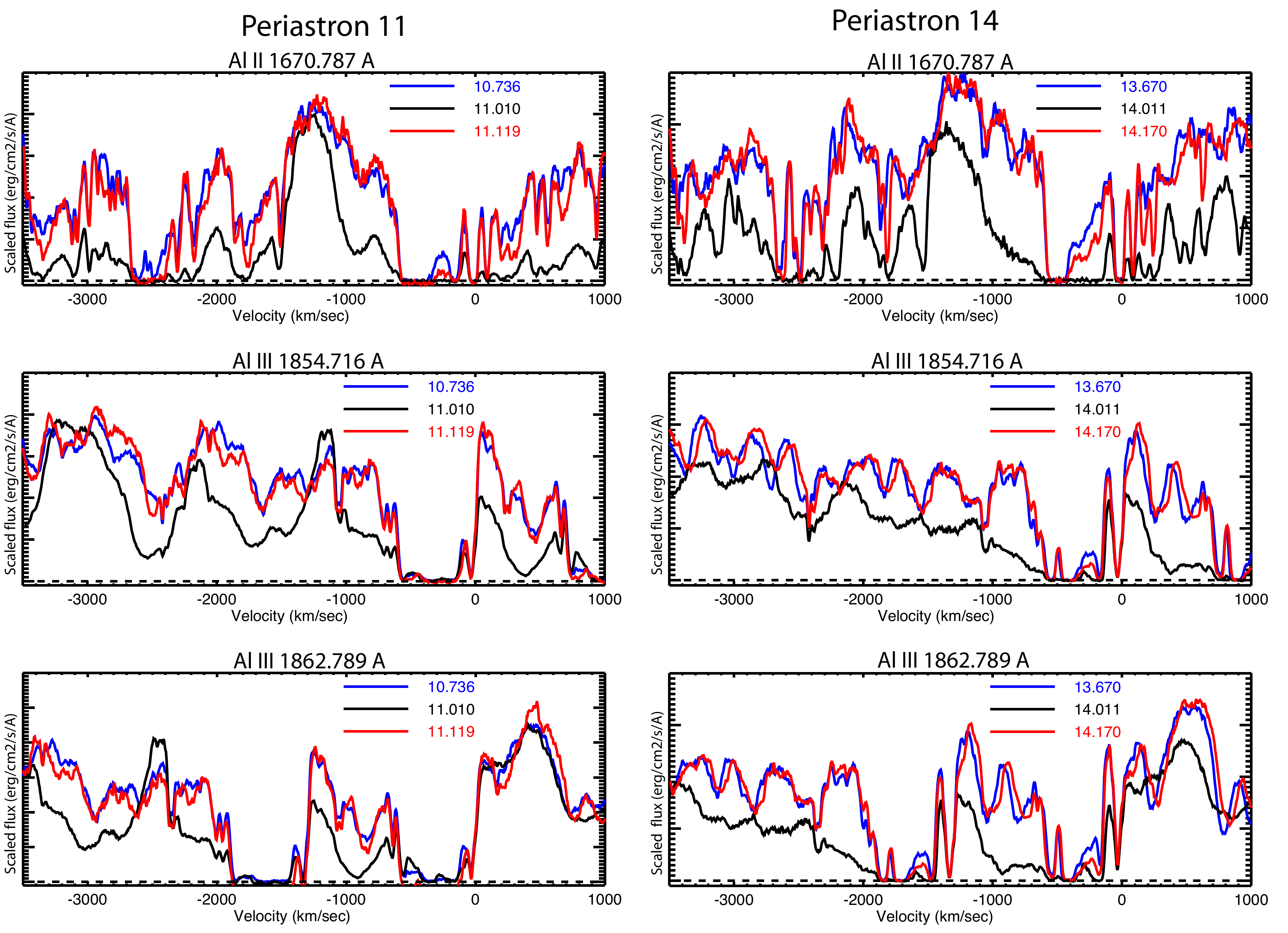}
\caption{Comparison of  \ion{Al}{2} and \ion{Al}{3} profiles across periastron passages 11 ({\bf Left Column}) and 14 ({\bf Right Column}). {\bf Top row:} \ion{Al}{2} $\lambda$1671.  Before and after periastron passage they are virtually identical. Across periastron passage 14, the ramp-like absorption extends from $-$60 to $-$1400 \kms\  and the $-$100 to $-$400 \kms\ region became saturated. {\bf Bottom two rows:} \ion{Al}{3} $\lambda\lambda$1855,1863 profiles. The \ion{Al}{3} $\lambda$1855 indicates possible continuous ramp absorption from $-$600 to $-$2200 \kms across periastron passage 14 while only to $-$1100 \kms across periastron passage 11.}
\label{fig:Al3Al2P}
\end{figure*}

Figures \ref{fig:C2C4N5P} to \ref{fig:Al3Al2P} compare profiles from well before periastron passage ($\phi =$ 10.736 and 13.670), early in the high-ionization recovery ($\phi =$ 11.119 and 14.170), and near periastron passage ($\phi =$ 11.010 and $\phi =$ 14.010) . Normalization was done using the ``continuum'' near 1483\AA\ for most lines except for \ion{N}{5}, which  was normalized using continuum longward of 1250\AA. Note in reference to the \ion{N}{5} velocity plots that the monotonic increase in flux from $-$3500 \kms\ conforms well to the wings of \ion{H}{1} Ly$\alpha$. As profiles for \ion{Al}{2} and  \ion{Al}{3} for all three epochs came from E230M spectra, the profiles in Figure \ref{fig:Al3Al2P} were scaled to ``continuum'' near 1800\AA.

\subsubsection{Changes across  Periastron passage 11}{\label{sec:Peri11change}}

\cite{Groh10} previously compared high velocity changes in seven resonant lines (excluding  \ion{N}{5}) leading up to periastron passage 11 (2003.5 to  high velocity changes  in the \ion{He}{1} $\lambda$10830 velocity profile leading up to the 2009.1 periastron. However, they did not discuss changes in the FUV near periastron passage 11. Here  we compare profiles before, during and after the periastron passages 11 and 14 from  $-4000<V< +1000$~\kms, with focus on changes near the periastron passage.

Profiles are plotted in velocity space for eight resonant lines in Figures  \ref{fig:C2C4N5P} through \ref{fig:Al3Al2P} (left columns) for three selected phases: high-ionization state before periastron passage ( {\color{blue}$\phi =$ 10.736}), low ionization state deep in periastron passage ($\phi =$ 11.010)  and early recovery to high-ionization state ({\color{red} $\phi =$ 11.119}). This comparison shows that changes in the  \ion{N}{5} profile are dominated by other absorptions at $\phi = 11.010$ (Figure \ref{fig:C2C4N5P}, lower left).  The \ion{C}{4} absorption increased in the $-$700 to $-$1200 \kms\ and possibly from $-$1500 to $-$1800 \kms\ ranges (Figure \ref{fig:C2C4N5P}, middle two left), but the \ion{C}{4} absorption in the $-$100 to $-$600 \kms\ range is obscured by iron absorptions as shown in Section \ref{sec:wind} and Figure \ref{fig:CMFGEN}.  Changes in the two \ion{Si}{4} velocity  profiles appear inconsistent because they are heavily modulated by lower ionization absorptions across periastron 11  (Figure \ref{fig:Si2Si4P}, lower two left). Increased absorption may be present from $-$1300 to $-$2200 \kms. 
The \ion{C}{2} velocity profile (Figure \ref{fig:C2C4N5P}, upper left) also appears to be modulated by other absorption lines at velocities more negative than $-$600 \kms.   On the other hand, the two \ion{Si}{2} profiles do not appear to be modulated by other absorption lines  (Figure \ref{fig:Si2Si4P}, upper two left). The \ion{Al}{2} velocity profile shows an increase in absorption from $-$600 to $-$1200 \kms\ near periastron passage (Figure \ref{fig:Al3Al2P}, upper left), while the two \ion{Al}{3} profiles show similar absorptions from $-$600 to$-$1200 \kms\ (Figure \ref{fig:Al3Al2P}, lower left two). The apparent absorptions extending above $-$3000 \kms\ in \ion{Al}{3} $\lambda$1855 are likely contributions from weaker absorbing lines.  The \ion {Si}{2} and \ion{Al}{2} absorptions saturate from $-$100 to $-$400 \kms at $\phi =$ 11.010 (Figure \ref{fig:Si2Si4P}, upper left two and Figure \ref{fig:Al3Al2P}, upper left).

 \subsubsection{Changes across Periastron passage 14}{\label{sec:Peri14change}}
 
Many  noticeable changes occur with the apparent brightening of \ec\ by periastron passage 14.  Velocity profiles of the same eight resonant lines are plotted in Figure  \ref{fig:C2C4N5P} through \ref{fig:Al3Al2P} (Right Columns) for three selected phases: high-ionization state before periastron passage ( {\color{blue}$\phi =$ 13.670}), low ionization state deep in periastron passage ($\phi =$ 14.010)  and early recovery to high-ionization state ({\color{red} $\phi =$ 14.170}).
We see that all resonant line profiles experienced increased absorptions across the interval from $-$600 to at least $-$1200 \kms\ (Figures \ref{fig:C2C4N5P} through \ref{fig:Al3Al2P}, right columns).  
The \ion{N}{5} resonant line absorption across $-$600 to $-$1400 \kms\ increased in absorption from {\color{blue} $\phi =$ 13.670} to $\phi =$ 14.011 then a peak in emission appears by {\color{red} $\phi =$14.170}.
The \ion{C}{4} absorptions at $\phi =$ 14.011 decreased from  $-$300 \kms\ continuously to $-$500 \kms\ relative to the profiles measured at {\color{blue} $\phi =$ 13.886} and {\color {red} $\phi =$ 14.170} (Figure \ref{fig:C2C4N5P}, right middle two). At $\phi =$ 14.010 the \ion{C}{4} absorption has dropped revealing the iron-dominated velocity profile of \ec-A (see Figure \ref{fig:CMFGEN}).
The two \ion{Si}{4} resonant lines  increased in absorption from $-$700  and $-$1800 \kms\ by $\phi =$14.011 relative to {\color{blue} $\phi =$ 13.667}. Both profiles then decreased in absorption by {\color{red} $\phi =$ 14.170}. 
Neither the \ion{C}{4} nor the \ion{N}{5} profiles are saturated in the 0 to $-$600 \kms\ range indicating that most of the primary wind is in a lower ionization state.
All four singly-ionized (\ion{Al}{2}, \ion{C}{2}, \ion{Si}{2} $\lambda\lambda$ 1527,1533) profiles reveal  very strong  increased absorption extending from $-$600 to $-$1500 \kms\ by $\phi =$ 14.011 relative to profiles measured at {\color{blue} $\phi =$ 13.670} and  return nearly to the pre-periastron profiles by {\color{red} $\phi =$ 14.011}.
The  resonant absorptions, extending from  $-$100 to $-$420 \kms\ of \ion{Si}{2} and \ion{Al}{2}, saturate by 
$\phi =$ 14.011, indicating absorbers most likely in the Homunculus and Little Homunculus relaxed from higher ionization states to singly-ionized state but returned to the high-ionization state by {\color{red} $\phi =$14.170}. This absorption was stronger across periastron passage 14 than across 11.

The most prominent changes are the  nearly linear absorption profiles ranging from $-$600 to $-$1500 \kms\ present in \ion{N}{5} through \ion{Al}{2}. This structure was noted by \cite{Gull21a} in singly-ionized absorption profiles. Likely  the transient absorption is the velocity-dispersed structure of the WWC trailing arm \citep{Madura12,Madura13}.

\subsubsection{Comparison of doublet profiles}{\label{sec:velocity}}

\begin{figure*}
\includegraphics[width=18cm]{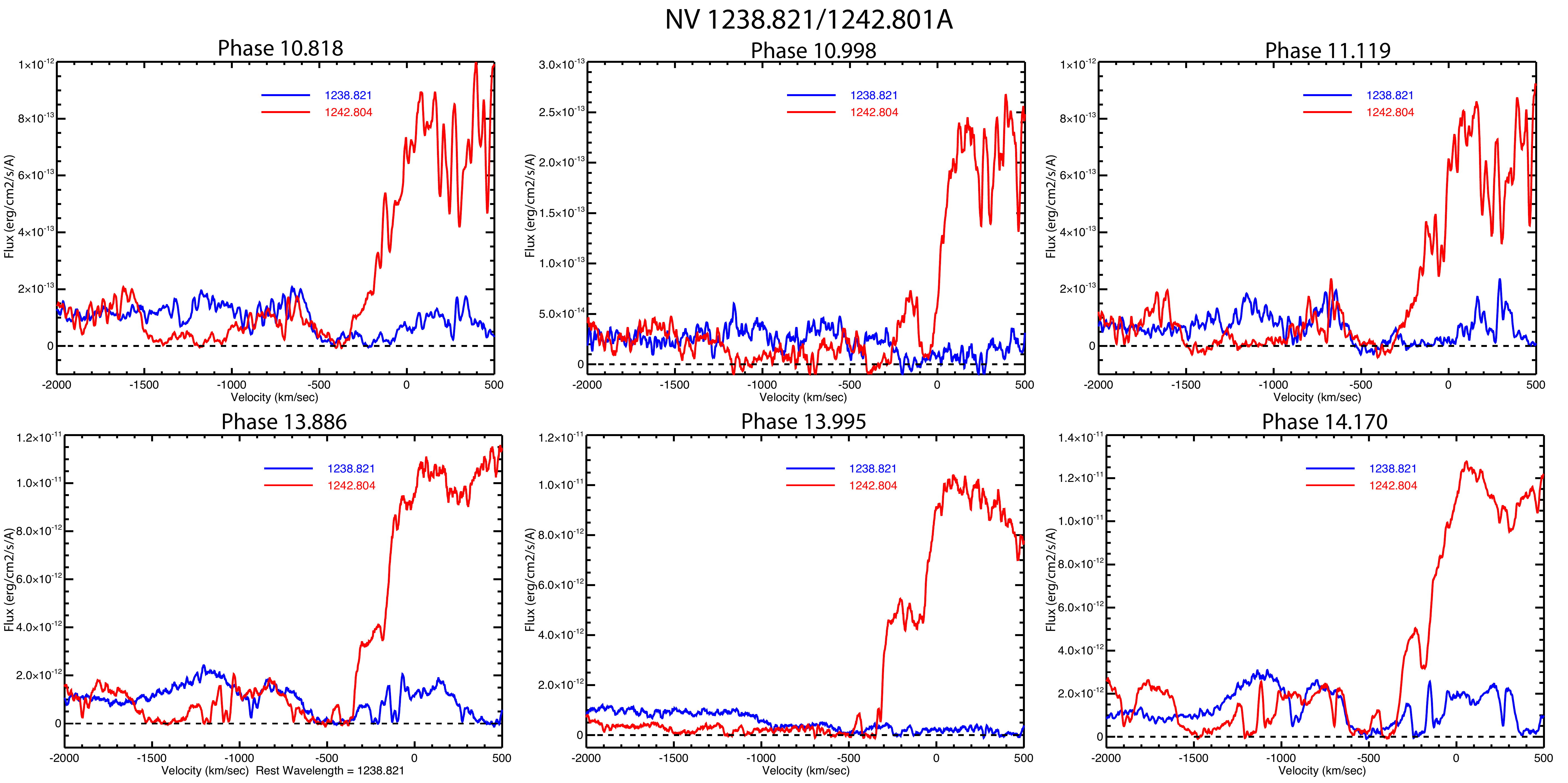}
\caption{Velocity profiles of  \ion{N}{5} $\lambda$1239,1243 doublet (IP $+$ 77.5 eV). The velocity comparison shows a broad absorption before and after,  but not during both periastron passages. {\bf Top Row:} For periastron passage 11, the $\approx$300 \kms-wide absorption is centered at $-$400 \kms. {\bf Bottom row:} periastron  14, the absorption has widened to $\approx$400 \kms\ and shifted to  $-$500 \kms. {\color{blue} The shorter wavelength is plotted in blue.} {\color{red}The longer wavelength is plotted in red.}}
\label{fig:NVPeri}
\end{figure*}

\begin{figure*}
\includegraphics[width=18cm]{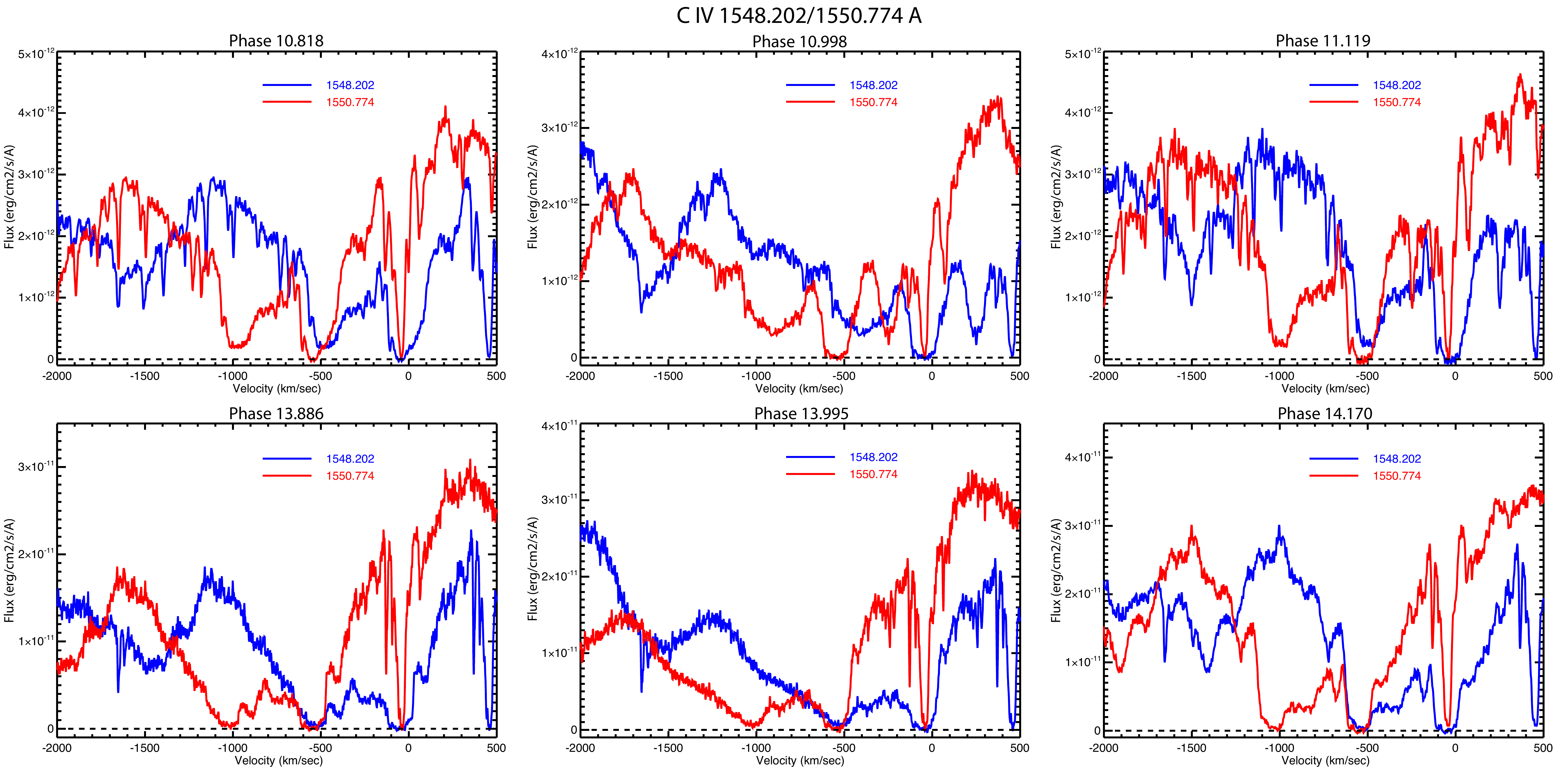}
\caption{Velocity profiles of the \ion{C}{4} $\lambda$1548,1551 doublet (IP $=$ 47.9 eV). {\bf Top Row:}  Reinforcement of profiles is limited. Before and after periastron passage 11, the velocity components appear to have been shifted in velocity but are confused by absorption from an iron line. The likely component is centered near $-$500 \kms and a few hundred \kms\ wide. Across periastron 11 ($\phi =$ 10.998), the profiles are dominated by iron lines with no \ion{C}{4} absorption present. {\bf Bottom row:} A common absorption component before and after periastron passage 14 is centered at $-$550 \kms with 200 \kms\ width and remains across periastron 14 ($\phi =$ 13.995) along with an absorption ramp that extends well beyond $-$800 \kms. {\color{blue} The shorter wavelength is plotted in blue.} {\color{red}The longer wavelength is plotted in red.}} 
\label{fig:CIVPeri}
\end{figure*}

\begin{figure*}
\includegraphics[width=17cm]{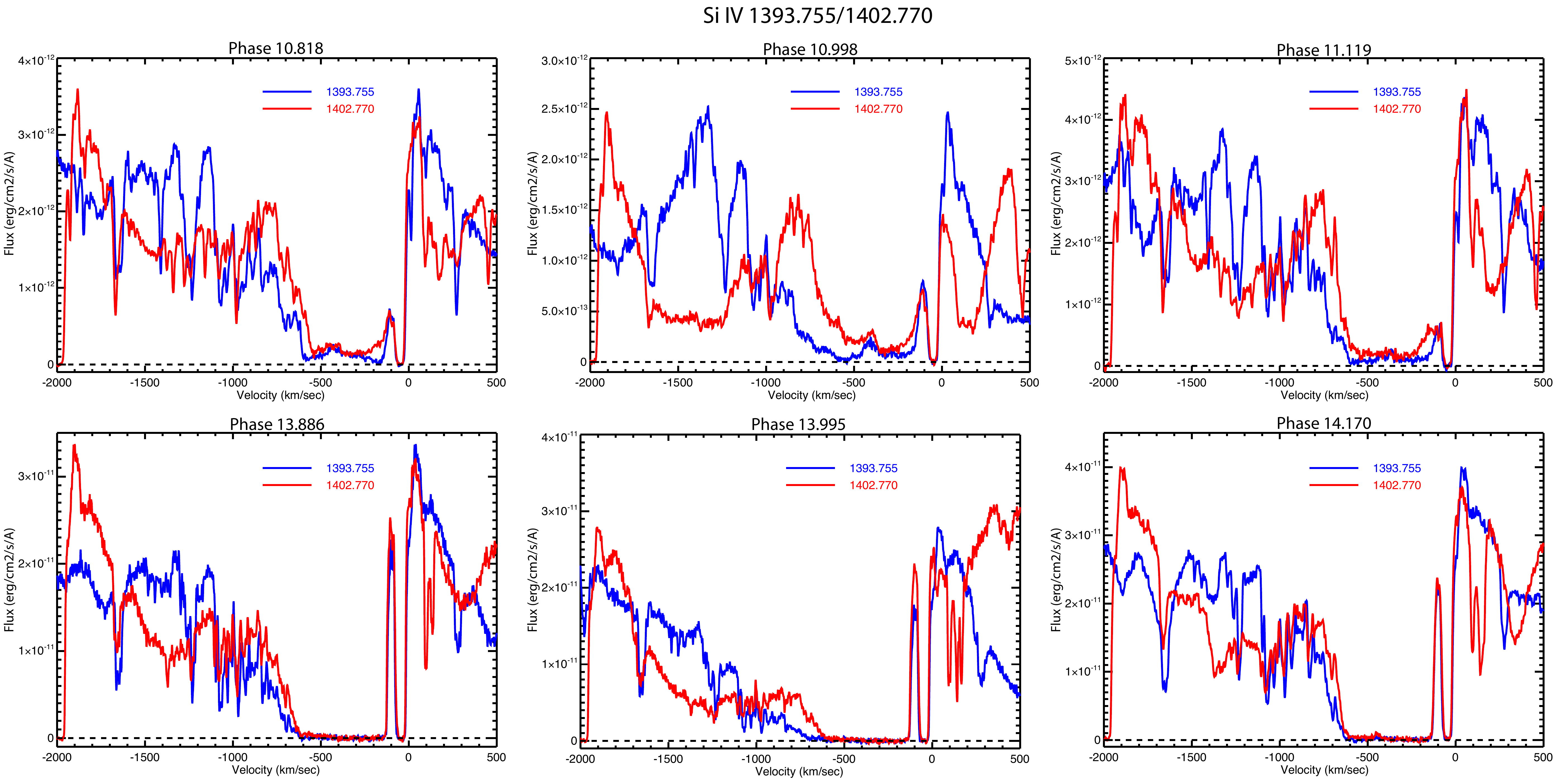}
\caption{Velocity profiles of the \ion{SI}{4} $\lambda\lambda$1394,1403 doublet (IP $=$ 33.5 eV). {\bf Top row: } The $-$100 to $-$600 \kms\ absorption, while strong, is not saturated throughout periastron passage 11. A strong absorption appears across periastron extending from $-$400 to $-$1000 \kms. {\bf Bottom row:} The $-$100 to $-$600 \kms\ velocity range is completely saturated before, during and after periastron passage 14. However, the high velocity profile does not track closely before and after the periastron passages.{\color{blue} The shorter wavelength is plotted in blue.} {\color{red}The longer wavelength is plotted in red.}}
\label{fig:SiIVPeri}
\end{figure*}

\begin{figure*}
\includegraphics[width=17cm]{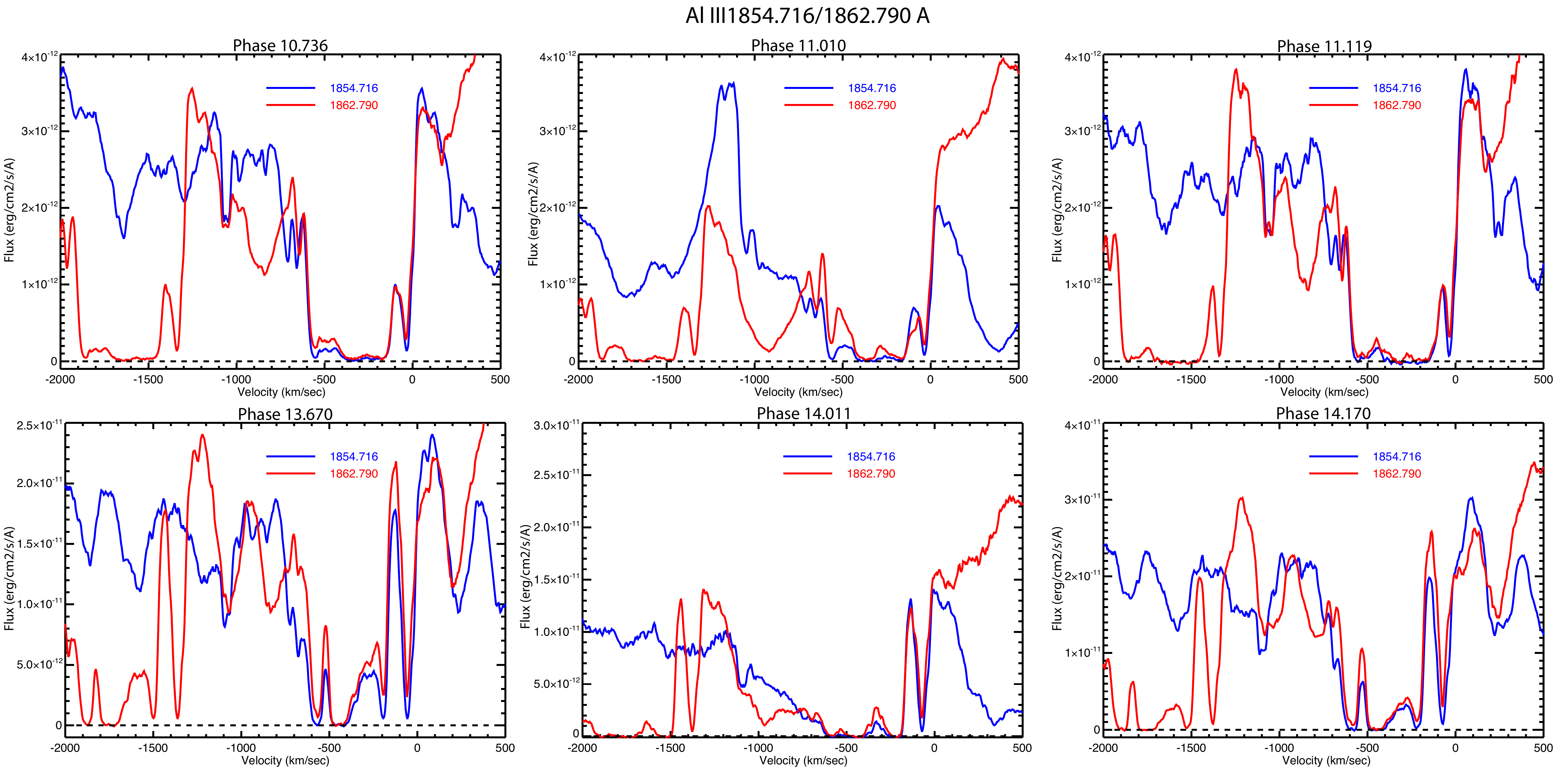}
\caption{Velocity profiles of the \ion{Al}{3} $\lambda\lambda$1855,1863 doublet (IP $=$ 28.5 eV). {\bf Top row:} Across periastron 11 ($\phi =$ 11.010), the $-$600 to $-$1200  \kms\ absorption increases. {\bf Bottom row:} The weak absorption islands at $-$100 and $-$300 \kms\ strengthen across periastron passage 14 ($\phi =$ 14.011) but return to pre-passage levels afterwards. The \ion{Al}{3} velocity profile was not available at $\phi =$ 13.995. {\color{blue} The shorter wavelength is plotted in blue.} {\color{red}The longer wavelength is plotted in red.}} 
\label{fig:AlIIIPeri}
\end{figure*}

\begin{figure*}
\includegraphics[width=17cm]{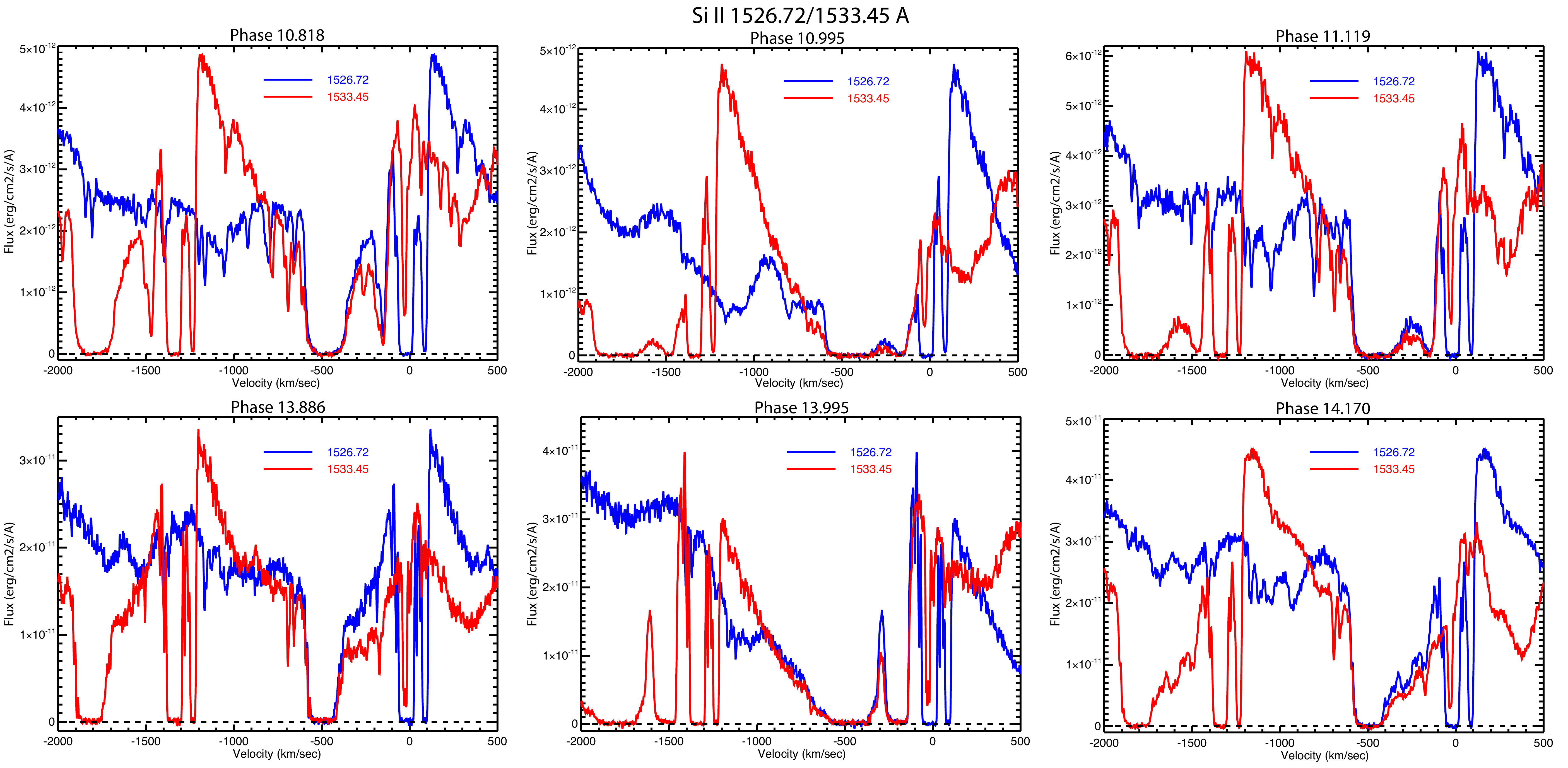}
\caption{Velocity profiles of the \ion{Si}{2} $\lambda\lambda$1527,1533 doublet (IP $=$ 8.2 eV). {\bf Top row:} Absorption  from $-$100 to $-$350 \kms\ is weak before, strengthens during,  and partially recovers after periastron passage 11. {\bf Bottom row:} Even weaker absorption is continuous from $-$100 to $-$400 \kms but saturated from $-$400 to $-$600 \kms before periastron. The ramp absorption occurs across periastron passage 14 ($\phi=$ 13.995) and disappears after periastron passage ($\phi =$ 14.170). The saturated absorption, while still centered close to $-$500\kms\ is narrower in width across periastron passage from 13 to 14. {\color{blue} The shorter wavelength is plotted in blue.} {\color{red}The longer wavelength is plotted in red.}} 
\label{fig:SiIIPeri}
\end{figure*}

Comparison of  doublet profiles  provides a critical test of the identification of  velocity components of the resonant lines, as we  noted in Section \ref{sec:doublet}. The doublet velocity comparison provides discrimination against absorptions from other lines. We apply this test to the same three spectra discussed in Section \ref{sec:Peri11change} and \ref{sec:Peri14change} to parse changes between the high-ionization and low-ionization states especially across the periastron passages (Figures \ref{fig:NVPeri} through \ref{fig:SiIIPeri}). The doublet velocity profile comparisons are not continuum normalized since the profiles are from the same spectrum.

The \ion{N}{5} $\lambda\lambda$1238, 1243  profiles were difficult to isolate  (Figure \ref{fig:NVPeri}). The increased density of absorbing lines below 1300\AA\ and the wings of the strong, extremely saturated \ion{H}{1} Ly$\alpha$\ profile depresses the continuum across the \ion{N}{5} velocity profile. The profiles presented in Figure \ref{fig:NVPeri} are smoothed to 20 \kms\ resolution to improve the signal to noise. 
The doublet profiles overlap in the high-ionization state before and after both periastron passages, but no evidence of \ion{N}{5} absorption is present close to periastron passage.  This suggests that \ec-B is the source of the ionization flux that producing N$^{+4}$ (IP $=$ 77.5 eV). \ion{N}{5} disappears when the ionizing flux from \ec-B is totally absorbed by the extended wind of \ec-A across periastron passage. 

Before and after periastron passage 11, the \ion{N}{5} absorption is centered at $-$400 \kms\ (Figure \ref{fig:NVPeri}, top row). The absorption width  at $\phi =$ 10.818  appears to increase from 250 \kms\  before passage to 400 \kms\ post-passage. Before and after periastron passage 14, the \ion{N}{5} absorption is centered at $-$450 \kms\ but does not appear to change significantly from 300 \kms\  in width (Figure \ref{fig:NVPeri}, bottom row).
The blueward absorption wing may be present in \ion{N}{5}  up to $-$1000 \kms\ at $\phi =$ 13.995, similar to  the high-velocity absorption  wing that extends from $-$600 to $-$1400 \kms\ in profiles from less-ionized lines, as noted above.  

The \ion{C}{4} $\lambda\lambda$1548,1552 doublet (IP $=$ 47.9 eV) is likewise complex due to confusing absorptions from the iron spectrum of \ec-A. The velocity overlap of the doublets is not completely consistent before, during and after periastron passage 11 (Figure \ref{fig:CIVPeri}, top row) as seen across periastron passage 14 (Figure \ref{fig:CIVPeri}, bottom row). Examination of the \cmfgen\ model profile (Figure \ref{fig:CMFGEN}, top right) indicates that this portion of the \ec-A  spectrum is dominated by iron absorptions. The velocity profile of  \ion{C}{4} $\lambda$1548  at $\phi =$ 10.998 (Figure \ref{fig:CIVPeri}, top row, center)  is quite similar to the absorbed profile generated by the \cmfgen\ model. Iron absorption  leads to  the less consistent overlap of the \ion{C}{4} profiles even in the high-ionization state before and after periastron passage 11. The \ion{C}{4} velocity profile in the high-ionization state appears to be unsaturated, centered at $-$450 \kms, with a  width of $\sim 100$~\kms.
In contrast, the profiles of the \ion{C}{4} doublet are consistent across periastron passage 14 (Figure \ref{fig:CIVPeri}, bottom row) most likely because of the greatly increased contribution in flux from the central core of \ec-A that led to a deeper, more direct view of the interacting wind structures  in our LOS.
The \ion{C}{4} profiles are centered at $-$550 \kms, and  saturated with a width of a few hundred \kms.
Comparison of variations in the \ion{C}{4} profiles across the two periastron passages reinforces the importance of the iron spectrum on this spectral interval. All profiles associated with periastron passage 11 show the influence of the iron spectrum in the 0 to $-$600 \kms\ interval (see Figure \ref{fig:CMFGENDoublet}). That contribution is  less apparent and reinforces the hypothesis that originally a small amount of FUV continuum  from the core of \ec-A passed through the occulter in our LOS.

The \ion{Si}{4} $\lambda\lambda$1394,1403 doublet is far more consistent than either the \ion{N}{5} or \ion{C}{4} doublets, likely because of its lower ionization energy (IP $=$ 33.5 eV). It characterizes structure more distant than the immediate WWC which produces the \ion{N}{5} and  \ion{C}{4} absorptions, where the ionizing flux has dropped in energy to well below  the IP of C$^{+3}$ (64.5 eV). The velocity overlaps are much better defined for both periastron passages (Figure \ref{fig:SiIVPeri}). 

None of the \ion{Si}{4} profiles near periastron passage 11 reach saturation even across the $-$100 to $-$600 \kms\ range.  This indicates that the LOS view of the interacting winds and shell structures are not completely ionized. Some decrease in absorption is notable at $\phi =$10.998 compared to just before and just after periastron  passage ($\phi =$ 10.818 and 11.119.  In contrast, that entire velocity range is saturated for all profiles associated with periastron passage 14. Along the LOS from the continuum-emitting core of \ec-A through the winds and the expanding shells, the continuum flux is entirely absorbed even deep in the low-ionization state induced by the periastron passage.

The \ion{Al}{3} $\lambda\lambda$1855,1863 doublet behaves very consistently across both periastron passages (Figure \ref{fig:AlIIIPeri}). Al$^{+2}$ has an IP $=$ 18.83 eV, well above the IP of hydrogen (IP $=$ 13.6 eV), while Al$^{+}$ has an IP = 5.99 eV. Changes in absorption by \ion{Al}{3} provide a reference for the ionization levels of the extended wind of \ec-A, the very cool wind-wind collision zones and the discrete absorbing shells within the Homunculus. 

The  \ion{Al}{3} absorption profiles associated with periastron passage 11 (Figure \ref{fig:AlIIIPeri}, top row) show only small 
changes 
between $-$100 and $-$ 600 \kms\ , which indicates very little change in the ionizing flux at 18.8 eV. In contrast, across periastron 14 (Figure \ref{fig:AlIIIPeri}, bottom row), significant changes occur in the low ionization state ($\phi = 13.995$ ) compared to phases before and after periastron passage. The absorption increased during the low ionization state, indicating that substantial amounts of aluminum must be in the high ionization state, Al$^{+3}$, which requires a source of $>$28.45 eV radiation. 
The \ion{Al}{3} profiles do not produce an as-well-defined strong component around $-$500 \kms\ as  do other resonant lines discussed in this paper.

The \ion{Si}{2} $\lambda\lambda$1527,1533 doublet shows the most variable changes before, during and after periastron passages (Figure \ref{fig:SiIIPeri}). Yet both doublet profiles track each other remarkably well. The \ion{Si}{2} absorption greatly increases across the low-ionization state, but the \ion{Si}{4} absorption remains saturated, indicating that much silicon is  doubly-ionized in the high-ionization state.

All high-ionization state profiles show a strong absorption component near $-$500 \kms. Before and after periastron passage 11, profiles (Figure \ref{fig:SiIIPeri}, top column left and right) show strong, saturated absorption centered at $-$475 \kms\ with a $200$~\kms\ width. Before and after periastron 14 profiles show a similar profile, centered at $-$500 km/s, saturated but narrower, with a $<$200 \kms\ width. \ion{Al}{3} is the exception as it shows a saturated absorption at $-$600 \kms\ even across periastron passage 14.

The $-$100 to $-$600 \kms\ velocity interval is strongly affected across the periastron passages but in very different ways in cycle 11 and 14. 
Before,  during and after periastron passage 11, a strong, relatively narrow, absorption centered at $-$150 \kms\ cuts the continuum, modulated by the  apparent P\,Cygni velocity profile, which persists to about $-$300 \kms\ and then drops to the strong absorption previously mentioned at $-$475 \kms. Very close to periastron passage ($\phi =$ 11.010), absorption becomes saturated, except for  weak emission between $-$200 and $-$300 \kms. By the early recovery to the high-ionization state ($\phi =$ 11.119) the absorption has partially decreased between $-$200 to $-$300~\kms.
In contrast, before periastron passage 14 ($\phi =$ 13.886), the P\,Cygni profile is recognizable  to $-$350\ \kms, especially for \ion{Si}{2} and \ion{Ni}{2} absorptions, and  the $-$150 \kms\ component is nearly absent. During the low ionization state near periastron 14, the P\,Cygni profile is nearly gone, as two saturated velocity components arise: $-$100 to $-$250 \kms\ and $-$325 to $-$ 600 \kms. By $\phi =$ 14.170, the continuum has recovered but only partially to the level seen in the P\,Cygni profile before periastron passage 14.

\section{Discussion}\label{sec:DIS}

In this paper we have made a detailed investigation into UV line profiles
as a function of cycle and orbital phase. While profile variations with orbital phase are linked with the highly eccentric orbit, the variations between
cycles 11 and 14 are most likely due to the disappearance of the
occulter, and not to an intrinsic variation in either of the stars.

Previously \cite{Davidson18} suggested that the brightening of \ec\ is caused by a decrease in the mass loss rate of \ec-A. However, the  X-ray light curve over the past five binary periods does not support such a claim  as  fluxes clocked by orbital phase  have not changed appreciably (except for a brief interval near the end of the X-ray minimum).  In particular, the X-ray flux measured at apastron is effectively the same from orbit to orbit, as is the rise of the light curve into periastron passage  \citep[][and Espinoza Galeas 2022, accepted]{espinoza21b}. Recovery of the post-periastron X-ray flux  does vary from passage to passage but is thought to be governed by instabilities of the WWC in  early recovery \citep{Corcoran17}. \cite{Pittard07} pointed out that 3-D hydrodynamic models of wind-wind shocked regions tend to smooth  and dissipate clumps as they enter the shocked structures. Hence, once the WWC zone has stabilized, variations across the extended high ionization state would be minimal. 

The systematic decrease in line equivalent widths (e.g., H$\alpha$, \ion{Fe}{2}) over the last decade could also be interpreted as evidence for stellar variability \citep{Mehner12, Mehner15}. However \citet{Damineli21} showed that the decreases could be better explained by the dissipation of the occulter. H$\alpha$, \ion{Fe}{2}, and [\ion{Fe}{2}] lines originate in the outer region of the wind, and were less affected by the  occulter. As the occulter dissipated (and/or moved out of the LOS), the EW would decline due to the rise in continuum flux. The existence of the occulter also explains why the models of \cite{Hillier06} underestimated the strength of H$\alpha$, and why H$\alpha$ EWs are lower when measured in spectra reflected from the lobes. The dissipating occulter now allows one to get a much better view of the WWC zone and  spectral changes near periastron.

We found further evidence for transient, high-velocity absorption wings seen in profiles near periastron 14, in  \ion{Si}{2}  \citep{Gull21a}  and other singly-ionized resonant lines extending across a range of ionization potential from 6 eV (\ion{Al}{2}) to  64.5 eV (\ion{C}{4}) (see Figures \ref{fig:C4C2} to \ref{fig:Al3Al2}). This high-velocity absorption wing occurs contemporaneously when the X-ray flux begins to drop, as the WWC possibly collapses for a short interval near periastron passage \citep{Hamaguchi07a,Madura13,Clementel15}.

\begin{figure*}
\includegraphics[width=17cm]{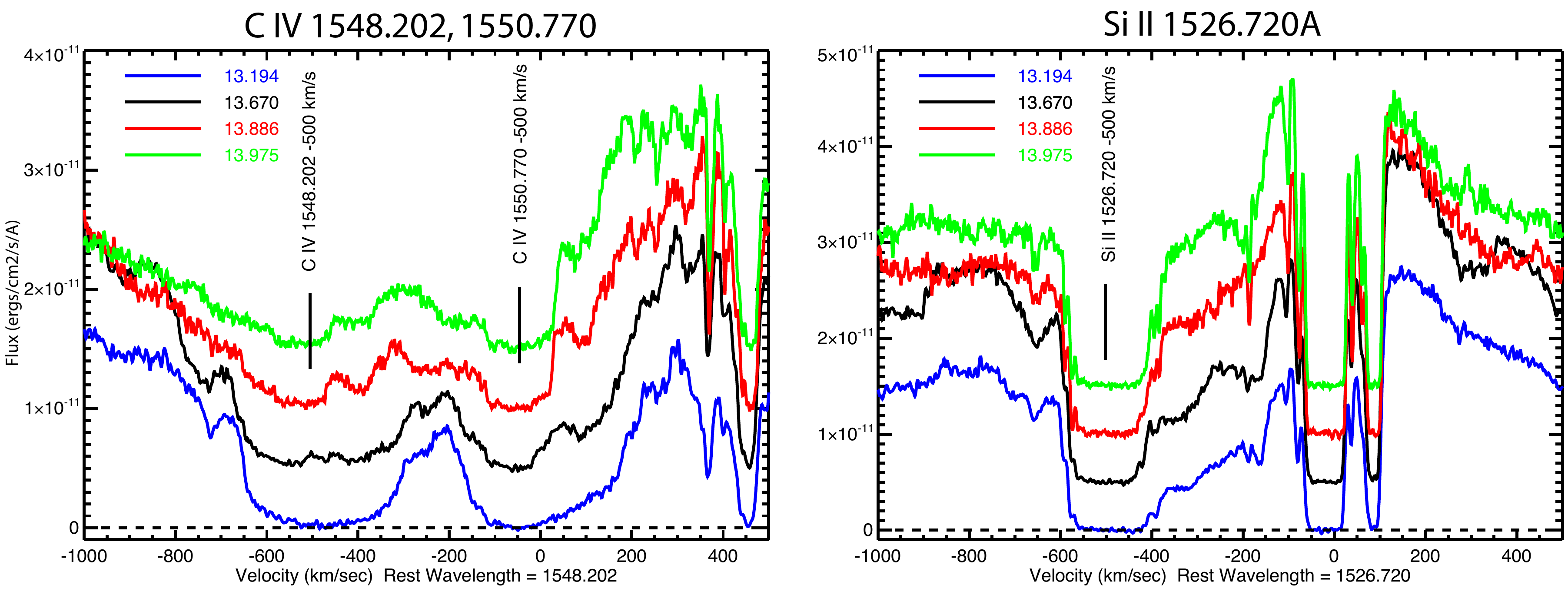}
\caption{{\bf Evidence of the evolving distant WWC in our LOS} Across cycle 13 (from 2014.6 to 2020.1) the \ion{C}{4} absorptions centered at $-$500 \kms\ have evolved from a broad $\approx$350 \kms-wide profile  at $\phi =$ 13.194 to a substantially weaker  $\approx$150 \kms-wide profile at $\phi =$13.975. Only a portion of the absorptions are black (total) at $\phi =$13.194. The \ion{Si}{2} $\lambda$1527 (and \ion{Al}{2}$\lambda$1671, not shown) absorption is saturated from $\approx -$425 to $\approx -$575 \kms\ with minimal changes across the entire cycle.} 
\label{fig:shift}
\end{figure*}

The downstream bowshock is the source of the $-$500 \kms\ absorption  seen in the resonant lines from \ion{Al}{2} to \ion{N}{5}. The primary wind that flowed for at least several weeks in our LOS across the periastron passage (extending from the trailing arm of the old WWC to the leading arm of the new WWC) is  shocked and accelerated by the secondary wind. Evidence of ionization and acceleration of this primary wind by the secondary wind is seen directly by the transient high velocity absorption wing near  periastron passage. The STIS FUV observations across periastron 14 show that this high-velocity absorption appeared by $\phi =$ 13.995 and persisted to $\phi =$ 14.011 (about 33 days). The wing is not present at $\phi =$ 13.975 and has disappeared by $\phi =$14.022, which places an upper limit of 94 days on its appearance.  This  wing extends redward of $-$550  to $-$1500 \kms\ and is present in most resonant line profiles (the \ion{N}{5} velocity profile was too noisy to confirm the presence of the high-velocity wing across this interval). 

By the early recovery to the high-ionization state, three spectra from early in cycles 11, 13, and 14  show strong absorptions at $-$500 \kms, but with very different widths (see Section \ref{sec:early} and Figure \ref{fig:Long}). 3D hydro models show that the downstream bowshock structure can attain a terminal velocity  up to V$_\infty=$ 1000 \kms\ in the orbital plane. We measure  V$_\infty \approx$ 500 \kms\ in our LOS which views the orbital plane at 49\degr. The 3D hydro models of \cite{Madura13} suggest that the half-opening angle of the wind-blown cavity is $\approx$45\degr\ and the observational evidence is that we view the binary from within the cavity. Hence the terminal velocity of the downstream shock should be somewhat more than the primary wind terminal velocity accelerated somewhat by the secondary wind and radiation pressure.  

3D hydro models show that a downstream bowshock persists for about 700 days, then breaks up \citep{Madura13, Madura14}. However, absorption at $-$500 \kms\ persists throughout the high ionization state (Figures \ref{fig:C4C2} through \ref{fig:Si2}) and across the entire cycle 13 (Figure \ref{fig:shift}). The \ion{C}{4}\,$\lambda\lambda$1458,1551 absorptions approach saturation only near $-$500 \kms\ at $\phi =$13.194, and become weaker with increasing orbital phase (Figure \ref{fig:shift}, left). However the lower-ionization resonant lines, like \ion{Si}{2} (Figure \ref{fig:shift}, right) are saturated from $-$425 to $-$575 \kms\ and change minimally across cycle 13. Further modeling and observations are needed to determine if this phenomenon repeats and/or evolves.

Across periastron passage 11, because of the occulter our LOS did not view much of the central core of \ec-A or  the effects of  \ec-B on the WWC and the inner wind. As \cite{Gull21a} demonstrated, despite similar samplings of spectra in orbital phases, the borehole effect was not detectable in the STIS FUV across periastron passage 11 as the continuum component originating from the core of \ec-A was mostly blocked by the occulter. The occulter dissipated by cycle~14, allowing the borehole effect to be detected in the FUV. Likewise, the bulk of the changes in resonant lines across periastron passage 11 were limited to changes  in the extended wind of \ec-A, but by periastron passage 14, modulation of the resonant lines by the WWC became much more obvious. 

Our LOS is quite unique and has helped parse out much new information on changes in the previously ejected shells and in the WWC. Moreover, these observations, combined with our understanding of how the ejecta is expanding and dissipating with time, gives us pause to consider how \ec\ might appear at later times. Namely, when the Homunculus and Little Homunculus have dissipated in thousands of years, then observers will have a clear view of the interacting binary if the binary members have not evolved to their final evolutionary stages. However the Homunculus and the Little Homunculus continue to expand and will drop in surface brightness, eventually merging into the ISM.
\section{Conclusions}{\label{sec:SUM}}

The most significant finding of this study of FUV resonant lines is  that we are now able to observe changes in the central core of the binary system and sample the effects of \ec-B and the WWC in our LOS. We summarize the main findings of this study below.

\begin {itemize}
\item The FUV flux has increased tenfold over the past two decades.
\begin {itemize} 
\item Most signatures of singly-ionized metals at velocities $-$121 to $-$168 \kms\ have disappeared, indicating that the Little Homunculus has become more highly ionized in our line of sight.
\item $\sim800$  absorption lines of H$_2$ within the Homunculus have disappeared  due photo-dissociation through the Lyman  bands.
\item NaD absorption components at $-$146 and $-$168 \kms\ have weakened or disappeared in the high-ionization state, but briefly return during the low-ionization state associated with  periastron passage (Pickett et al., in prep.)
\item Absorption in the $-$100 to $-$400 \kms\ velocity range has decreased in \ion{C}{2}, \ion{Si}{2}, \ion{Al}{2} and \ion{Al}{3} resonant lines,  while absorption increased in \ion{Si}{4}.
\end{itemize}
\item The bore hole, created by the passage of \ec-B through the extended wind of \ec-A near periastron passage, became visible across periastron 14, but was not seen  across periastron passage 11 \citep{Gull21a} because the  occulter preferentially blocked the continuum from the core of \ec-A relative to emission from  its extended wind. 
\item The H$\alpha$ emission equivalent width continues to decrease as the visible flux increases with the decreasing obscuration of the central core. However, slowing of the decline in  H$\alpha$  equivalent width \citep{Damineli21} may indicate that the rate of brightening is decreasing. 
\item  The high-velocity absorption component of \ion{C}{4} appears to have increased up to $-$2600 \kms across the high state in cycle 13 relative to cycle 10 (Figure \ref{fig:C4C2}, left). 
\item A transient absorption wing  extending sfrom  $-$600 \kms\ to  $-$1500~\kms, appeared across periastron passage 14, but was not apparent across periastron passage 11. Its character is astonishingly similar across all resonant lines. We do not know if this is a one-time variation or a long-term trend, but suspect the latter. The most likely explanation is the trailing arm of the WWC passed through the LOS between the \ec-A continuum core in our LOS and our vantage point.
\item With the decreasing absorption of singly-ionized metals in the multiple shells of the Little Homunculus, the  profiles  of \ion{Si}{2}, \ion{Ni}{2} and \ion{Al}{2}  become more consistent with profiles generated by \cmfgen\ models of \ec-A.
\item  The decrease in optical depth of the occulter allows us to used  FUV resonant lines to provide a means to track the effects of the interacting winds and directly monitor any changes in flux and spectral features. The influence of the Homunculus absorbing systems in the LOS is decreasing and will continue to decrease.
\item We find no significant evidence for changes in the winds of the binary members.
\end{itemize}

Future studies, especially in the FUV, become all the more important. Hopefully \hst/STIS access will continue through the next decade as monitoring of changes of the FUV resonant absorptions before and during  the next periastron passage in 2025 is now key to confirming the changes seen across periastron passage 14, assuming the changes in the winds are repeating.

The challenge posed by the FUV resonant lines must be met by  modeling of the massive binary system and its wind interactions plus the changes in the multiple shells within the Homunculus and the Little Homunculus.

\begin{acknowledgments} NR acknowledges funding from \hst\ programs 15611 and 15992 which were accepted as supplementary observations associated with CHANDRA programs 20200564 and 21200197. MFC is supported under the CRESST-II cooperative agreement \#80GSFC17M0002 with the NASA/Goddard Space Flight Center. AD acknowledges FAPESP for support through process 2011/51680-6. AJFM is grateful for financial aid from NSERC (Canada). The work of FN is supported by NOIRLab, which is managed by the Association of Universities for Research in Astronomy (AURA) under a cooperative agreement with the National Science Foundation. CMPR acknowledges support from the National Science Foundation under Grant No. AST-1747658. TRG received no direct support for this study. We thank the referee for the careful and very prompt review of this lengthy study.
\end{acknowledgments}
\facilities{HST(STIS)}
\software{\cmfgen\      \citep{Hillier01a,Hillier11} }
\bibliography{ref}{}
\bibliographystyle{aasjournal}
\end{document}